\documentclass[prb,twocolumn,amsmath,amssymb,superscriptaddress,floatfix,nofootinbib]{revtex4-2}
\usepackage{epsfig, graphicx,graphics,amsmath,amssymb,float}
\usepackage[T1]{fontenc}
\usepackage[latin9]{inputenc}
\usepackage{amsmath}
\usepackage{amssymb}
\usepackage{appendix}
\usepackage{amscd}
\usepackage{bm}
\usepackage{psfrag}
\usepackage{bbm} 
\usepackage{babel}
\usepackage{wasysym }
\usepackage{mathrsfs}
\usepackage{color}
\usepackage[normalem]{ulem}

\begin{document}

\title{Dynamical parity selection in superconducting weak links}

\author{Nico Ackermann}
\affiliation{
Departamento de F{\'i}sica Te{\'o}rica de la Materia Condensada, Condensed Matter Physics Center (IFIMAC)
and Instituto Nicol{\'a}s Cabrera, 
Universidad Aut{\'o}noma de Madrid, 28049 Madrid, Spain}

\author{Alex Zazunov}
\affiliation{Institut f\"ur Theoretische Physik, Heinrich-Heine-Universit\"at, D-40225  D\"usseldorf, Germany}

\author{Sunghun Park}
\affiliation{
Departamento de F{\'i}sica Te{\'o}rica de la Materia Condensada, Condensed Matter Physics Center (IFIMAC)
and Instituto Nicol{\'a}s Cabrera, 
Universidad Aut{\'o}noma de Madrid, 28049 Madrid, Spain}

\author{Reinhold Egger}
\affiliation{Institut f\"ur Theoretische Physik, Heinrich-Heine-Universit\"at, D-40225  D\"usseldorf, Germany}

\author{Alfredo Levy Yeyati}
\affiliation{
Departamento de F{\'i}sica Te{\'o}rica de la Materia Condensada, Condensed Matter Physics Center (IFIMAC)
and Instituto Nicol{\'a}s Cabrera, 
Universidad Aut{\'o}noma de Madrid, 28049 Madrid, Spain}

\begin{abstract}
Excess quasiparticles play a crucial role in superconducting quantum devices ranging from qubits to quantum sensors. In this work we analyze their dynamics for phase-biased finite-length weak links  
with several Andreev subgap states, where the coupling to a microwave resonator allows for 
parity state (even/odd) readout.  Our theory shows that almost perfect dynamical polarization in a given parity sector is achievable by applying a microwave pulse matching a transition in the opposite parity sector. 
Our results qualitatively explain key features of recent experiments on hybrid semiconducting nanowire 
Josephson junctions and provide theoretical guidelines for efficiently controlling the parity state of Andreev qubits.
\end{abstract}
\maketitle

\section{Introduction}

Superconducting Josephson devices are basal ingredients 
for many of the currently most advanced platforms for quantum information processing 
and quantum-limited measurements \cite{Xiang2013,Wendin2017,Kjaergaard2020,Blais2021,Rasmussen2021}. In such devices, non-equilibrium quasiparticles have been identified as one of the main sources of relaxation and decoherence \cite{Martinis2009}, and different techniques have been proposed for mitigating their effect \cite{Corcoles2011,Wang2014,Marin-Suarez2020}. On the other hand,  
recent fabrication progress on few-channel hybrid nanowires, e.g., semiconducting InAs wires with a superconducting Al shell, has resulted in the wide availability of high-quality tunable superconducting weak links \cite{Zgirski2011,Bretheau2013,Janvier2015,Larsen2015,Lange2015,Higginbotham2015,Woerkom2017,Hays2018,Tosi2019,Hays2020,Whiticar2021,Hays2021,Fatemi2022,Matute2022,Wesdorp2022}, harboring Andreev bound states (ABSs) with phase-dependent energy  below the pairing gap ($|\varepsilon|<\Delta$) 
\cite{Beenakker1991,Furusaki1991,Bagwell1992,Nazarov2009}. A qubit can be encoded by ABSs \cite{Desposito2001,Zazunov2003,Nazarov2003,Zazunov2005,Padurariu2010}  if the fermion number parity of the Andreev sector (referred to as ``parity'' below) is conserved. In practice, coherent qubit operation is limited to times shorter 
than the parity switching time $\tau_{\rm p}$ caused, e.g., by
transitions between ABSs and above-gap continuum levels. 
For even-parity \cite{Janvier2015,Hays2018} as well as odd-parity \cite{Tosi2019,Hays2020,Hays2021} Andreev qubits, 
coherent qubit manipulation has already been demonstrated on time scales much smaller than
$\tau_{\rm p}\sim 100~\mu$s.
It stands to reason that reaching a thorough understanding of, and thereby good control over, the 
parity dynamics in superconducting weak links is of fundamental and applied importance. 

A remarkable recent experiment~\cite{Wesdorp2022} has achieved long parity lifetimes together with 
almost complete dynamical parity polarization.  An initial microwave pulse is tuned to induce
transitions to an excited many-body Andreev state within the even (or odd) parity sector.  
After the pulse, deterministic and close-to-ideal parity polarization in the \emph{opposite} odd (even) parity sector has been observed for times $t\alt \tau_{\rm p}$, while the steady-state ABS populations are recovered only for $t\gg \tau_{\rm p}$ \cite{Wesdorp2022}.
We here present a unified microscopic theory which qualitatively explains basic experimental observations 
and can be used as starting point for a more detailed description including, e.g., spin-orbit coupling and electron-electron interaction effects.  We show that for the dynamical parity 
selection to occur, weak links of intermediate length (harboring at least two ABSs) are needed.   
While the case of short weak links (with a single ABS) is well understood \cite{Kos2013,Zazunov2014,Olivares2014,Riwar2015,Park2020}, our theory 
describes the parity dynamics in superconducting weak links of arbitrary length.
Focusing on the simplest nontrivial case with two ABSs, we arrive at an intuitive physical picture which identifies wide parameter regimes suitable for dynamical parity selection, and thus for coherent qubit manipulation.  

\begin{figure}[ht!]
\includegraphics[scale=0.25]{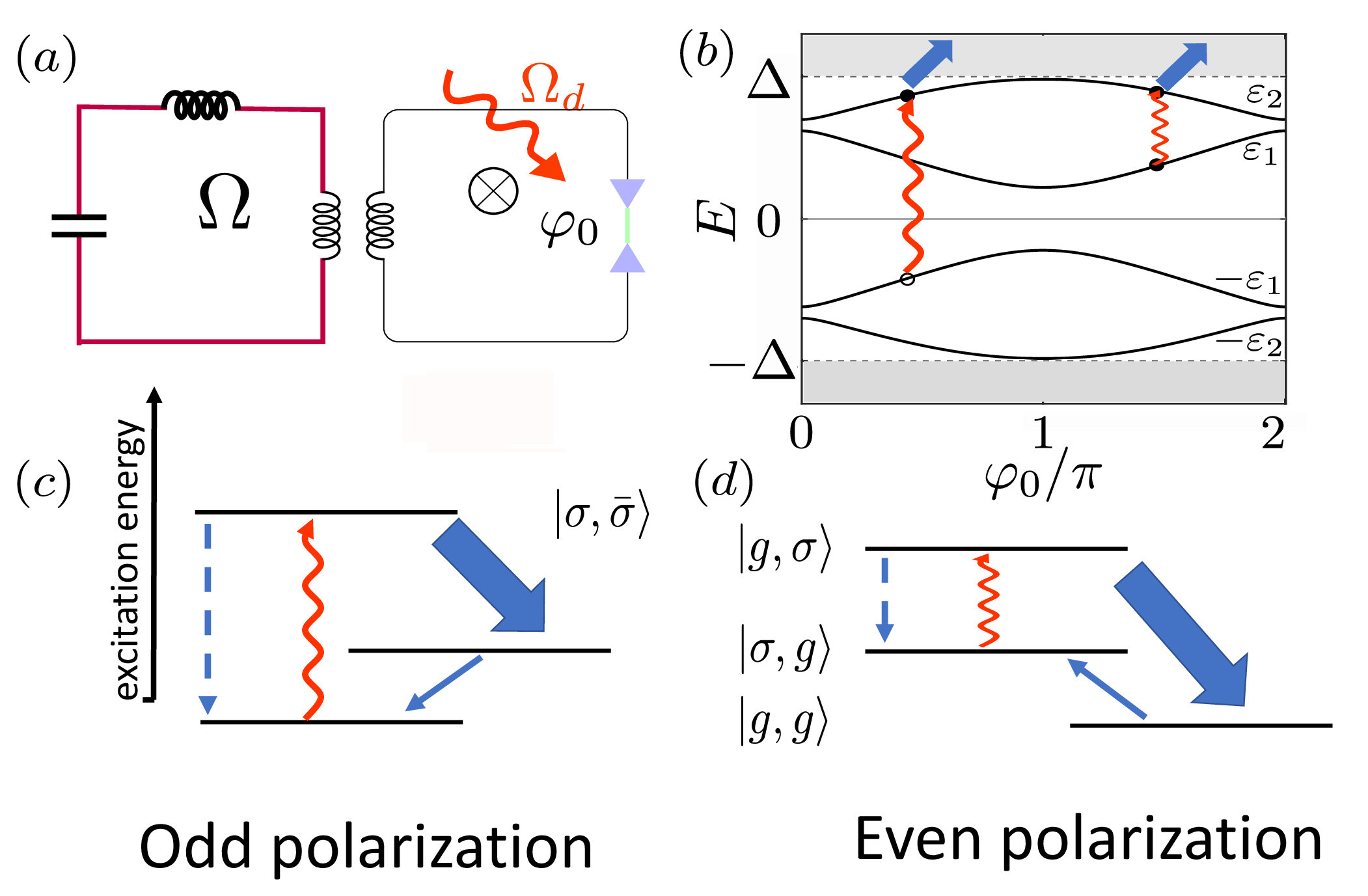}
\caption{
(a) Schematic setup: a loop containing a finite-length Josephson junction is inductively coupled to a microwave resonator with resonance frequency $\Omega$. The flux on the loop imposes the average phase bias $\varphi_0$ on the junction. In all panels, red wavy lines indicate a microwave pulse with drive frequency $\Omega_d$. 
Solid (dashed) blue arrows show parity-changing (parity-conserving) transitions. 
The associated rates determine the arrow thickness.
(b) ABS energies $E=\pm\varepsilon_{1,2}$ vs $\varphi_0$ for junction 
length parameter $\ell=1.3$ and transmission probability ${\cal T}=0.76$, 
see Eq.~(\ref{ALdispeq}). 
Thick arrows indicate dominant parity-changing transitions due to resonator photon absorption.
Panels (c) and (d) show the relevant many-body Andreev states, see Table \ref{tab1}, 
and transition rates for achieving dynamical parity polarization:  (c) Excitation of the 
mixed pair transition $|g,g\rangle \rightarrow |\sigma,\bar \sigma\rangle$ induces odd-parity polarization.
(d) Excitation of the odd-parity transition $|\sigma, g\rangle \rightarrow |g,\sigma\rangle$ 
leads to even-parity polarization.}
\label{fig1}
\end{figure}

Before entering a detailed theoretical discussion, let us briefly summarize the basic picture.
As schematically shown in Fig.~\ref{fig1}(a),
we consider a single-channel superconducting weak link embedded in a loop geometry and coupled to an LC circuit. The average phase difference $\varphi_0$ is related to the magnetic flux 
threading the loop. The weak link has the length $L=\ell \xi_0$, with the coherence length $\xi_0=v_F/\Delta$ and the Fermi velocity $v_F$ (we put $\hbar=e=k_B=1$). 
For a short link with $\ell\ll 1$, it is well known that a single ABS with particle-hole symmetric energy levels $\pm \varepsilon(\varphi_0)$ exists \cite{Nazarov2009}, where $0\le \varepsilon(\varphi_0) <\Delta$. In the semiconductor picture \cite{Zazunov2014}, there are four many-body Andreev states at fixed $\varphi_0$:
(i) for the ground state $|g\rangle$ with $E_g=0$, taken as the reference energy for fixed $\varphi_0$, only the $-\varepsilon$ level is occupied; (ii)  
for the excited state $|e\rangle$ with $E_e=2\varepsilon$ and the same (even \cite{Zazunov2014}) parity, only the $+\varepsilon$ level is occupied; (iii) for the odd-parity states $|\sigma=\pm \rangle$ with $E_\sigma=\varepsilon$, the $\pm\varepsilon$ levels are either both empty or both occupied. On the other hand, for a link of intermediate length $\ell\sim 1$, one typically finds 
\emph{two} ABSs with $0\le\varepsilon_1(\varphi_0)\le \varepsilon_2(\varphi_0)<\Delta$, as illustrated in Fig.~\ref{fig1}(b). The resulting 16 many-body Andreev states are specified 
in Table~\ref{tab1}.  Compared to the short-junction limit, longer junctions harboring at least two ABSs allow for qualitatively new phenomena such as dynamical parity polarization.  Our theory addresses precisely this situation.

In Fig.~\ref{fig1}(c), we show the relevant states for generating
odd-parity polarization together with the dominant transition rates due to photon-induced mixing of
ABSs and continuum states.  Here the initial pulse is assumed resonant with the 
$|g,g\rangle\leftrightarrow|\sigma,\bar\sigma\rangle$ transition, 
where $\sigma'=\bar\sigma=-\sigma$ in Table \ref{tab1} because of spin conservation. 
The drive frequency is thus given by $\Omega_d=\varepsilon_1+\varepsilon_2$. If 
spin-orbit and Coulomb interactions are neglected, such transitions are degenerate 
with the $|\sigma,g\rangle\leftrightarrow|e,\sigma\rangle$ transition in the odd sector, 
which could then be excited simultaneously. 
However, even mild interactions are known to remove such degeneracies 
\cite{Matute2022}, see Sec.~\ref{sec3b} for details, and one expects that single transitions 
can be driven in practice.
If the excited state is connected to the lowest-energy odd-parity state
by a parity-changing transition rate $\Gamma_{|\sigma,\bar\sigma\rangle\to |\sigma, g \rangle}$ exceeding the relaxation rate into the even-parity ground state, 
the system is quickly driven into the odd-parity sector with almost ideal polarization.
The much smaller rate $\Gamma_{|\sigma,g\rangle\to |g, g\rangle}$ then only 
establishes the final steady-state population reached at very long times 
$t\gg\tau_{\rm p}$.   
A similar picture applies for the dynamical stabilization of 
the even-parity sector, see Fig.~\ref{fig1}(d).
Our formalism allows us to identify optimal parameters for achieving dynamical parity selection. 
For instance, a key criterion is to have the upper ABS manifold with energy 
$\varepsilon_2(\varphi_0)$ within an energy range $\sim \Omega$ near the continuum threshold $\varepsilon=\Delta$, 
together with temperatures comparable to the resonator frequency $\Omega$. Under these conditions,
 a thermally activated photon is available, photon absorption becomes effective, and parity-changing transitions are favored over parity-conserving ones.

\begin{table}[t!]
\centering
\begin{tabular}{| c | c | c|| c | c | c|}
 \hline
 $|\alpha,\beta\rangle$ & $E_{\alpha,\beta}$ & Parity & $|\alpha,\beta\rangle$ & $E_{\alpha,\beta}$ & Parity \\ \hline
 $|g,g\rangle$ & $0$ & even & $|\sigma, g\rangle$ & $\varepsilon_1$ & odd \\
 $|e,g\rangle$ & $2\varepsilon_1$ & & $|g, \sigma\rangle$ &  $\varepsilon_2$ & \\
 $|\sigma, \sigma'\rangle$ & $\varepsilon_1+\varepsilon_2$ & &  $|e,\sigma\rangle$ & $2\varepsilon_1+\varepsilon_2$ & \\
 $|g,e\rangle$ & $2\varepsilon_2$ & & $|\sigma ,e\rangle$ & $\varepsilon_1+2\varepsilon_2$ & \\
 $|e,e\rangle$ & $2\varepsilon_1+2\varepsilon_2$ & & & &\\ 
 \hline 
\end{tabular}
\caption{Many-body Andreev states $|\alpha,\beta\rangle$ with energy $E_{\alpha,\beta}$ for a weak link with two ABSs at fixed $\varphi_0$, see Fig.~\ref{fig1}(b). The first (second) index $\alpha~(\beta)~\in \{ g,e,\sigma=\pm\}$ refers to 
the occupation of the ABS pair with energy $\pm \varepsilon_1$ ($\pm \varepsilon_2$).}
\label{tab1}
\end{table}

The structure of the remainder of this article is as follows.  We discuss the model underlying our study 
and the corresponding dynamical equations in Sec.~\ref{sec2}.  Dynamical polarization effects are 
addressed in detail in Sec.~\ref{sec3}.   The paper concludes with a summary in Sec.~\ref{conc}.
Technical details about our derivations can be found in the Appendix.

\section{Model and master equation}\label{sec2}

In this section, we describe the model underlying our study (cf.~Sec.~\ref{sec2a}) and the resulting quantum dynamics 
in the Andreev sector (cf.~Sec.~\ref{sec2b}).  Before studying dynamical parity polarization effects in Sec.~\ref{sec3},
we summarize our parameter choices and the steady-state solution in Sec.~\ref{sec2c}.

\subsection{Model}\label{sec2a}

We study a phase-biased superconducting weak link given by a nanowire Josephson junction embedded in a loop geometry and coupled 
to the electromagnetic phase fluctuations of an LC circuit, see Fig.~\ref{fig1}(a). 
In the standard Andreev approximation \cite{Nazarov2009}, the Hamiltonian $H_{L/R}$ for the $s$-wave BCS superconductor on the left ($x<0$) or right ($x>0$) side of the contact is expressed by quasiclassical Nambu spinor envelopes $\psi_\pm(x,t)$ for right- and left-movers with Fermi momentum $\pm k_F$, respectively. 
With $s_{L/R}=\pm 1$ and Pauli matrices $\tau_{x,y,z}$ in Nambu space, and
assuming the same pairing gap $\Delta$ on both sides, one finds  
\cite{Zazunov2005,Nazarov2009,Zazunov2014}
\begin{eqnarray} \nonumber
    H_{j=L/R}(t)& =& \sum_{\pm} \int_{s_j x < 0} dx \,  \psi^\dagger_\pm(x,t) \bigl ( 
\mp i  v_F  \tau_z \partial_x \\ &+& V_j(t) \tau_z + \Delta \tau_x e^{i \tau_z \phi_j(t)}
\bigr)\psi^{}_\pm(x,t), \label{HLR}
\end{eqnarray}
where the fluctuating voltages $V_{j}(t)$ and phases $\phi_j(t)$ are linked by the Josephson relation,
$V_j=\dot \phi_j/2$, and the gauge-invariant phase difference across the contact is $\varphi(t)=\phi_L-\phi_R$. 
Modeling the weak link as a normal-conducting constriction with length $L=\ell \xi_0$ 
and transmission probability ${\cal T}$ due a local scatterer, 
the quasiclassical envelopes on both sides of the contact $(x=0^\pm)$ are matched by 
a transfer matrix \cite{Zazunov2005,Nazarov2009,Zazunov2014},
\begin{equation}
\sqrt{\cal T} \left( \begin{array}{l} \psi_+(0^-,t) \\ \psi_-(0^-,t) \end{array} \right) =
\left( \begin{array}{cc} e^{-i \tau_z \hat \theta_R} & r  \\ r
& e^{i \tau_z \hat \theta_R} \end{array} \right)
\left( \begin{array}{l} \psi_+(0^+,t) \\ \psi_-(0^+,t) \end{array} \right)
\label{BC}
\end{equation}
with the reflection amplitude $r=\sqrt{1-{\cal T}}$ and the dynamical 
phase shift $\hat \theta_R(t)=\hat \theta(t)-\frac{L}{2v_F}\dot \phi_R(t)\tau_z$, where
$\hat \theta=\frac{L}{v_F} i\partial_t=\ell \omega/\Delta$.
Equation \eqref{BC} assumes a symmetric contact with a local scatterer 
at the center of the weak link. The generalization to arbitrary impurity position is 
straightforward, cf.~App.~\ref{appA}. 
Since our conclusions are robust when changing this position,  we focus on the symmetric case.

We next write $\varphi(t)=\varphi_0+\tilde\varphi(t)$ with $|\tilde\varphi(t)|\ll 1$, where  $\tilde\varphi(t)$ describes electromagnetic phase fluctuations. As described in detail in App.~\ref{appA},
after choosing the gauge $\phi_{L/R}(t)=\pm \varphi(t)/2$,  we proceed 
by expanding $H_j(t)$ to linear order in $\tilde\varphi$, gauging away 
$\dot{\tilde\varphi}$ from the matching condition \eqref{BC}, and expanding to linear order also
in $\dot{\tilde\varphi}$.  Finally, the phase difference $\varphi_0$ is gauged away from $H_j(t)$ 
such that $\varphi_0$ appears only through the transfer matrix. 
For $\tilde\varphi=0$, one arrives at the analytically solvable Bogoliubov-de~Gennes (BdG) equations
\begin{eqnarray}
&& \left(-i v_F  \tau_z \sigma_z\partial_x + \Delta \tau_x \right) \Psi_\nu(x) = \varepsilon_\nu \Psi_\nu(x),
\label{BdG}\\ \nonumber
&&\sqrt{\cal T} \Psi_\nu(0^-) = e^{i \tau_z \varphi_0/2}
\left( e^{-i \tau_z \sigma_z \theta_\nu} + r \sigma_x \right)\Psi_\nu(0^+),
\end{eqnarray}
where the stationary eigenstates $\Psi_\nu(x)$ for energy $\varepsilon_\nu$ are bispinors 
in Nambu and left-right mover space,  
 Pauli matrices $\sigma_{x,y,z}$ act in left-right mover space, and $\theta_\nu= \ell \varepsilon_\nu/\Delta$. The solution of Eq.~\eqref{BdG} defines the noninteracting Hamiltonian $H_0$ and is given in 
 App.~\ref{appA}. The quantum numbers $\nu$ include subgap ABS solutions, $\nu=\pm m$ with $m=1,\ldots,N$ and
$\varepsilon_{\nu}= \pm \varepsilon_m(\varphi_0)$, where we find the condition 
\begin{equation}\label{ALdispeq}
\cos^{-1}(\varepsilon/\Delta) - \ell\varepsilon/\Delta \pm \sin^{-1} \left( \sqrt{\cal T} \, 
\sin(\varphi_0/2) \right) = \pi m.
\end{equation}
An example with $N=2$ ABS pairs is shown in Fig.~\ref{fig1}(b).
Above-gap continuum quasiparticles are instead labeled by $\nu=(\varepsilon,s)$ with $|\varepsilon|>\Delta$ 
and $s\in \{1,2,3,4\}$ for an electron or hole incoming from the left or right side.  

We then employ a second-quantized description where the 
field operator, $\hat \Psi(x)=\sum_\nu \Psi_\nu(x) \gamma_\nu$, is expressed in terms of 
fermion annihilation operators $\gamma_\nu$.
The noninteracting Hamiltonian is 
\begin{equation}\label{H0def}
H_0=\sum_\nu \varepsilon_\nu \hat{n}_\nu,\qquad \hat{n}_\nu=\gamma_\nu^\dagger \gamma_\nu^{}.
\end{equation}
From the above steps, we find  $H_c = \frac12 {\cal I}\tilde \varphi$ for the coupling to 
phase fluctuations, where the supercurrent operator ${\cal I}$ has the matrix elements  
\begin{equation}\label{calI}
{\cal I}_{\nu\nu'} = \int dx \, {\rm sgn}(-x) \Psi^\dagger_{\nu}  
\left(\frac{\omega_{\nu\nu'}}{2 i} 
(\tau_z -\ell \tau_y\sigma_z) +  \Delta \tau_y \right) \Psi_{\nu'}^{} 
\end{equation}
with $\omega_{\nu\nu'}=\varepsilon_{\nu}-\varepsilon_{\nu'}$.
The full Hamiltonian, $H=H_0+H_c+H_{\rm bath}$, also includes an oscillator bath term 
for the electromagnetic environment \cite{Nazarov2009,Weiss2012,BreuerPetruccione}, for which we assume a thermal equilibrium state at temperature $T_{\rm env}$.
For the setup in Fig.~\ref{fig1}(a), with resonance frequency $\Omega$, dimensionless coupling strength $\kappa$, and damping constant $\eta$, the bath spectral density is taken as \cite{Weiss2012,Zazunov2014} 
\begin{equation}\label{specdens}
    J(\omega) = \frac{\kappa^2\eta}{2\pi}
    \left( \frac{1}{(\omega - \Omega)^2 +\frac{\eta^2}{4}}
    -\frac{1}{(\omega + \Omega)^2 +\frac{\eta^2}{4}} \right)+J_{\rm ohm}.
\end{equation}
For a realistic comparison with experimental results, we here also include a background Ohmic spectral 
density, $J_{\rm ohm}(\omega)=\alpha_0 \omega e^{-|\omega|/\omega_c}$   \cite{Weiss2012,BreuerPetruccione},
with dimensionless coupling $\alpha_0\ll 1$ and high-frequency cutoff $\omega_c\approx \Delta$. 

\subsection{Master equation}\label{sec2b}

We next turn to the time-dependent density matrix $\rho(t)$ describing 
the fermion sector, which follows after tracing over the bath. 
Starting from $H = H_{0}+H_c+H_{\rm bath}$,  the density matrix of the complete (system-plus-bath) 
system in the interaction picture,
$\tilde \rho_{\rm tot}(t)$, obeys the Liouville-von Neumann equation, 
\begin{eqnarray}\label{eomrhotot}
\dot{\tilde \rho}_{\rm tot}(t) &=& - i \left[ H_c(t), \tilde \rho_{\rm tot}(t) \right],
\\ \nonumber
H_c(t) &= &\frac{\tilde \varphi(t)}{2} \sum_{\nu,\nu'} 
e^{i \omega_{\nu\nu'} t} {\cal I}_{\nu\nu'} \gamma_\nu^\dagger \gamma^{}_{\nu'}.
\end{eqnarray} 
Integrating Eq.~\eqref{eomrhotot} and inserting the result back into Eq.~\eqref{eomrhotot}, we obtain
\begin{eqnarray}\nonumber
\dot{\tilde \rho}_{\rm tot} &=& - \int_0^t d \tau
\left( H_c(t) \left[ H_c(t-\tau), \tilde \rho_{\rm tot}(t-\tau) \right]+ {\rm h.c.} \right) \\ &-& i \left[ H_c(t), \tilde \rho_{\rm tot}(0) \right].
\end{eqnarray}
Following Refs.~\cite{Weiss2012,BreuerPetruccione}, we assume that the bath always remains 
in thermal equilibrium, $\tilde \rho_{\rm tot}(t) \approx \tilde \rho(t)
\otimes \rho_{\rm bath}$, that the system-bath coupling is small 
(Born approximation), and that the bath has a very short memory time (Markov approximation). 
These conditions are met for temperatures large against all microscopic transition rates, which for our system parameters implies 
$T_{\rm env}\agt 10^{-2}\Delta$, and for weak system-bath coupling $\kappa\ll 1$. 
The reduced density matrix $\tilde \rho(t)$ for the fermionic sector is then obtained by tracing over the bath, 
$\tilde \rho(t) = {\rm Tr}_{\rm bath} \, \tilde \rho_{\rm tot} (t)$, resulting in
\begin{equation}
\dot{\tilde \rho} = \int_0^\infty d \tau\,  \mathfrak{D}(\tau)\left[ {\cal I}(t - \tau) \tilde \rho(t), {\cal I}(t)  \right] +{\rm h.c.}
\end{equation}
with the bath correlation function
\begin{equation}
\mathfrak{D}(t) = {\rm Tr}_{\rm bath} \left(\rho_{\rm bath} \, \frac{\tilde \varphi(t)}{2} \,
\frac{\tilde \varphi(0)}{2} \right) = \mathfrak{D}^\ast(-t).
\label{Dtau}
\end{equation}
Writing $\mathfrak{D}(\tau) = \int_{-\infty}^\infty \frac{d \omega}{2 \pi} \, e^{-i \omega \tau} \mathfrak{D}_\omega$
and introducing a bath spectral density $J(\omega)=-J(-\omega)$, we can express 
$\mathfrak{D}_\omega$ as 
\begin{equation}\label{Dom}
\mathfrak{D}_\omega = 2 \pi J(\omega) \left( n_B(\omega) + 1 \right),
\end{equation}
with $n_B(\omega)=1/(e^{\omega/T_{\rm env}}-1)$.  For the setup in Fig.~\ref{fig1}(a),
$J(\omega)$ is given by   Eq.~(\ref{specdens}).

Collecting the above results and switching to the Heisenberg picture, we arrive 
at a standard Lindblad equation \cite{Weiss2012,BreuerPetruccione}
\begin{equation}\label{Lindbleq2}
\dot \rho = - i [ H_0, \rho ] +\sum_{\nu\ne\nu'} \Gamma_{\nu \nu'} {\cal L}(Q_{\nu'\nu}) \rho ,
\end{equation}
with ${\cal L}(Q)\rho= Q \rho Q^\dagger -\frac12 \left\{ Q^\dagger Q, \rho  \right\}$.
For the quasiparticle transition $\nu \rightarrow \nu'$, the
jump operator $Q_{\nu'\nu}=\gamma_{\nu'}^\dagger \gamma_{\nu}^{}$ comes with the transition rate
\begin{equation}\label{Gammanumu}
\Gamma_{\nu \nu'} = 2\pi \left| {\cal I}_{\nu' \nu} \right|^2 \, J(\omega_{\nu \nu'}) \left[n_B(\omega_{\nu\nu'})+1\right].
\end{equation}
With the notation $\bar\nu$ for labeling the
opposite-energy partner obtained by particle-hole symmetry, $\varepsilon_{\bar \nu}=-\varepsilon_\nu$,
the rates \eqref{Gammanumu} satisfy the symmetry relation 
$\Gamma_{\bar \nu'\bar\nu} = \Gamma_{\nu\nu'}.$
In addition, forward and backward rates are linked by detailed balance,
\begin{equation}
\Gamma_{\nu\nu'} = e^{\omega_{\nu\nu'} / T_{\rm env}} \, \Gamma_{\nu'\nu}.
\end{equation}
In general, the Lindblad equation also contains an additional Lamb shift, $H_L$, 
such that $H_{0}\to H_{0}+H_L$ in Eq.~\eqref{Lindbleq2}. The Lamb shift describes a renormalization of the 
quasiparticle energy levels. With Eq.~\eqref{Dom}, we find
\begin{eqnarray}
H_L& =& \sum_{\nu\ne \nu'} Y(\omega_{\nu'\nu}) \left| {\cal I}_{\nu\nu'} \right|^2 Q_{\nu\nu'}^\dagger Q_{\nu\nu'}^{},\\ \nonumber
Y(\omega)& =& {\rm p.v.} \int_{-\infty}^\infty \frac{d\omega'}{2 \pi} \, \frac{\mathfrak{D}_{\omega'}}{\omega - \omega'},
\end{eqnarray}
where `p.v.' indicates a principal value integration.
We note that $H_L$ commutes with $H_{0}$. In fact, $H_L$
does not enter the matrix rate equations \eqref{dotP} discussed below at all, and we therefore ignore Lamb shift contributions in what follows.

Following Ref.~\cite{Zazunov2014}, we further simplify Eq.~(\ref{Lindbleq2}) by neglecting entanglement 
between Andreev and continuum states, i.e., $\rho(t)\simeq \rho_A(t)\otimes\rho_c$ with the reduced density matrix $\rho_A(t)$ for the Andreev sector. Using
a distribution function $n_{p=(\varepsilon,s)}$ for parametrizing the continuum states, we write
\begin{equation}
\rho_c = \prod_{p} \left( n_p \left| 1_p \rangle \langle 1_p \right| + \left( 1 - n_p \right) \left| 0_p \rangle \langle 0_p \right|\right),
\end{equation}
with $|1_p\rangle=\gamma_p^\dagger|0_p\rangle$.  We can then
trace Eq.~(\ref{Lindbleq2}) over the continuum states.
As a consequence,  $\rho_A(t)$ obeys a simpler Lindblad equation with source terms describing transitions from or into the 
continuum sector,  
\begin{eqnarray}\label{Lindbleq3}
\dot \rho_A &=& -i[H_0,\rho_A]+\sum_{\lambda\ne \lambda'} 
\Gamma_{\lambda\lambda'} {\cal L}(Q_{\lambda'\lambda}) \rho_A \\
\nonumber &-&\sum_\lambda\Gamma^{\rm out}_{\lambda}\left(\frac12\left\{\hat{n}_\lambda,\rho_A\right\}-
\gamma_\lambda^{} \rho_A \gamma^{\dagger}_\lambda\right)  \\ \nonumber
&-&\sum_{\lambda}\Gamma^{\rm in}_\lambda 
\left(\frac12\left\{1-\hat{n}_\lambda,\rho_A\right\}-\gamma^{\dagger}_\lambda \rho_A \gamma_\lambda^{} \right) ,
\end{eqnarray}
with $\hat n_\lambda$ in Eq.~\eqref{H0def} and the rates 
\begin{equation}\label{gammainout}
\Gamma^{\rm in}_\lambda= \sum_p \Gamma_{p\lambda} n_p, \quad 
\Gamma^{\rm out}_\lambda= \sum_p \Gamma_{\lambda p} \left(1-n_p\right),
\end{equation} 
which describe transitions from the continuum into the Andreev sector and vice versa.
We model the presence of excess continuum quasiparticles by a quasi-equilibrium Fermi distribution,
\begin{equation}\label{npqp}
n_{p=(\varepsilon,s)}= \frac{1}{e^{\varepsilon/T_{\rm qp}}+1},
\end{equation}
where the quasiparticle temperature $T_{\rm qp}$ may differ from (and typically will exceed) the temperature
$T_{\rm env}$ characterizing the electromagnetic environment. 

Focusing on the case $N=2$, we finally project Eq.~\eqref{Lindbleq3} into the 
many-body Andreev states $|\alpha,\beta\rangle$  listed in Table \ref{tab1}. 
The diagonal elements of $\rho_A$ contain the respective occupation probabilities, 
$P_{\alpha,\beta}(t) = \langle \alpha,\beta|\rho_A(t)|\alpha,\beta\rangle$, which are 
combined to form a 16-dimensional vector ${\bf P}(t)$.  
Since the dynamics of the off-diagonal part of $\rho_A(t)$ decouples from ${\bf P}(t)$,
we obtain a matrix rate equation for the occupation probabilities alone, 
\begin{equation}\label{dotP}
{\bf \dot{P}}(t) ={\bf M} \, {\bf P}(t),
\end{equation}
where the real symmetric $16\times 16$ matrix ${\bf M}$ is expressed in terms of the rates \eqref{Gammanumu} und \eqref{gammainout}.  
Additional simplifications are possible by exploiting state degeneracies in Table \ref{tab1}, 
e.g., $P_{+,g}=P_{-,g}$.  We discuss this issue and provide explicit expressions 
for ${\bf M}$ in App.~\ref{appB}. Given the eigenvalues $\lambda_n\le 0$ and the corresponding eigenvectors ${\bf P}_n$ of ${\bf M}$,
we arrive at the solution ${\bf P}(t) = \sum_{n} c_n e^{\lambda_n t} {\bf P}_n$, 
where the initial configuration ${\bf P}(0)$ is determined by the real coefficients $c_n$.
Given the time-dependent solution of Eq.~\eqref{dotP},
the parity-resolved total occupation probabilities are 
\begin{equation}
P_{\rm even}(t)=\sum_{\alpha,\beta\in \{g,e\}} P_{\alpha,\beta} + \sum_{\sigma,\sigma' \in {\pm}} P_{\sigma,\sigma'}
\end{equation}
and $P_{\rm odd}(t) = 1 - P_{\rm even}(t)$.

\subsection{Parameter choice and steady-state solution}\label{sec2c}
 
For the numerical results shown below, we consider a weak link
with reduced length $\ell=L/\xi_0=1.3$ and transmission probability ${\cal T}=0.76$, where the ABS spectrum is 
shown in Fig.~\ref{fig1}(b).  The parameters entering $J(\omega)$ in Eq.~\eqref{specdens} 
were chosen as 
\begin{equation}\label{parameters}
\Omega=0.13\Delta, \, \kappa=0.1, \, \eta=10^{-4}\Delta,\, \alpha_0=10^{-3}.
\end{equation}
This parameter choice is aimed at describing, at least qualitatively, the experimental situation in Ref.~\cite{Wesdorp2022}, where the resonator frequency 
was $\Omega\approx 4.82$~GHz (which roughly corresponds to $\Omega=0.13 \Delta$ for the case of Al) and
the quality factor was $Q  \approx 1.7 \times 10^3$ (which implies $\eta = \Omega/Q \approx 10^{-4} \Delta$). 
The Ohmic background constant $\alpha_0$ was chosen as a fit parameter \sout{in Ref.~\cite{Wesdorp2022}}.
While the parameter choice \eqref{parameters} is motivated by the experimental setup of Ref.~\cite{Wesdorp2022},
it is worth emphasizing that our general conclusions are robust against parameter changes, see Sec.~\ref{sec3}.

\begin{figure}
\includegraphics[width=0.8\columnwidth]{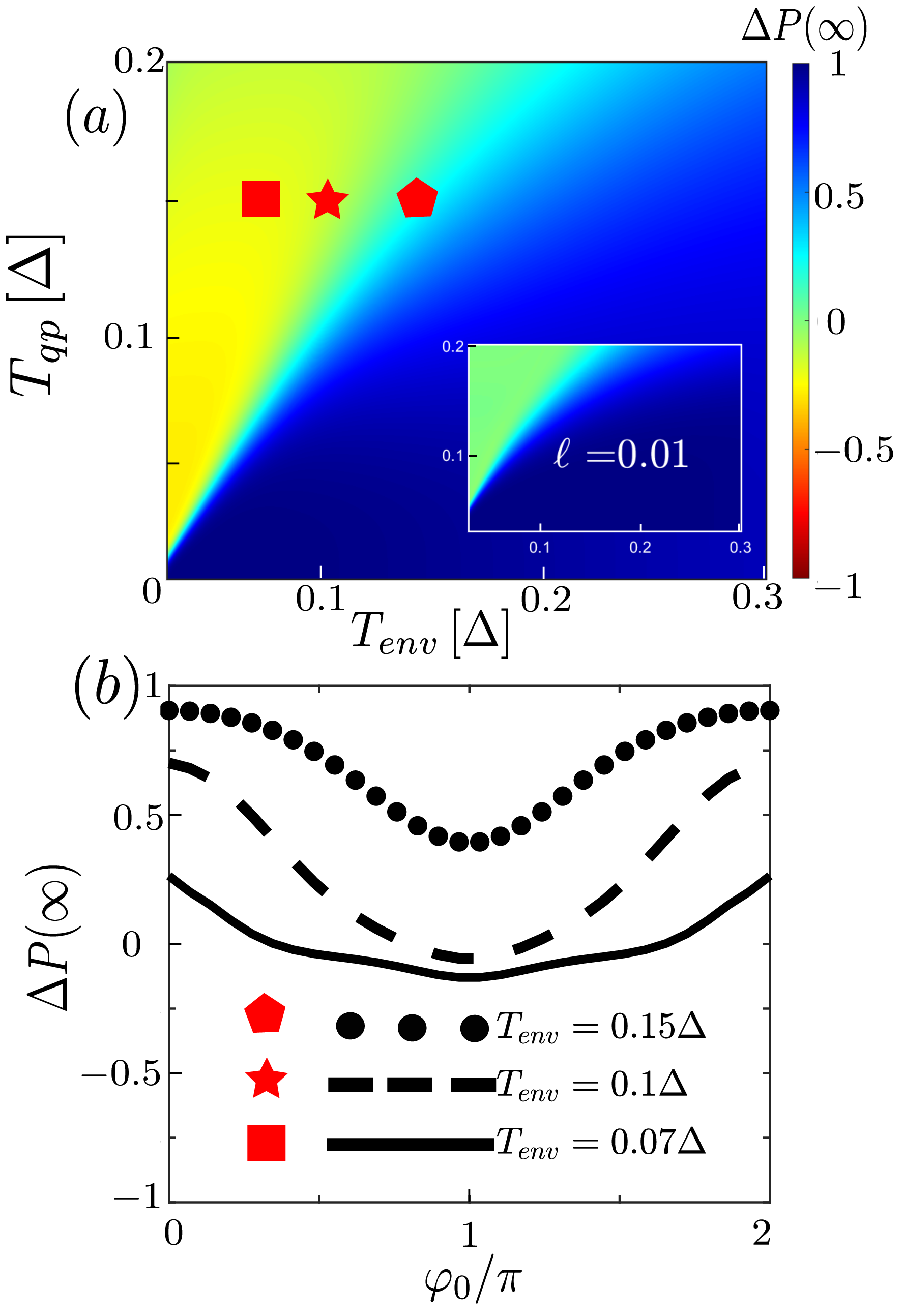}
\caption{Difference $\Delta P(\infty)=P_{\rm even}(\infty)-P_{\rm odd}(\infty)$ of the 
steady-state parity probabilities for $\ell=1.3$, ${\cal T}=0.76$, and the parameters \eqref{parameters}. 
(a) Color-scale plot of $\Delta P(\infty)$ in the $T_{\rm env}$-$T_{\rm qp}$ plane for $\varphi_0=\pi$. The inset shows the corresponding results for $\ell=0.01$ and otherwise identical parameters.  (b) $\Delta P(\infty)$ vs $\varphi_0$ for $T_{\rm qp}=0.15\Delta$ and three different values of $T_{\rm env}$.}  
\label{fig2}
\end{figure}

Let us close this section by addressing the steady state reached for long times $t\to \infty$. 
The steady-state solution of Eq.~\eqref{dotP}, ${\bf P}(\infty)$, follows from ${\bf M}\, {\bf P}(\infty) =  0$ 
together with  normalization, $\sum_{\alpha,\beta} P_{\alpha,\beta} = 1$. 
In Fig.~\ref{fig2}(a), we show results for the probability difference $\Delta P(\infty) \equiv P_{\rm even}(\infty)-P_{\rm odd}(\infty)$ as color-scale plot in the $T_{\rm env}$-$T_{\rm qp}$ plane for $\varphi_0=\pi$. 
We observe that the odd-parity sector becomes favorable for large ratio $T_{\rm qp}/T_{\rm env}$, which can be rationalized by noting that excess continuum quasiparticles can then proliferate,
see Fig.~\ref{fig2}(a).
In fact, for large $T_{\rm qp}/T_{\rm env}$,  the ratio 
\begin{equation}
   \frac{\Gamma^{\rm in}_{\lambda=1}}{\Gamma^{\rm out}_{\lambda=1}} 
\sim e^{(\Delta-\varepsilon_1)/T_{\rm env}} n_p(\Delta)
\end{equation} 
can exceed unity, which turns out be be a necessary condition for population inversion of the steady-state occupations. 
Although much more suppressed, such a population inversion can also occur in short weak links with only a single ABS pair, see the inset of Fig.~\ref{fig2}(a).  
We emphasize that the steady state typically does not allow for a large parity polarization in the odd-parity sector. 
As we illustrate in Fig.~\ref{fig2}(b), this only partial steady-state parity polarization in addition 
strongly depends on the phase difference $\varphi_0$.  

\begin{figure}
\includegraphics[scale=0.25]{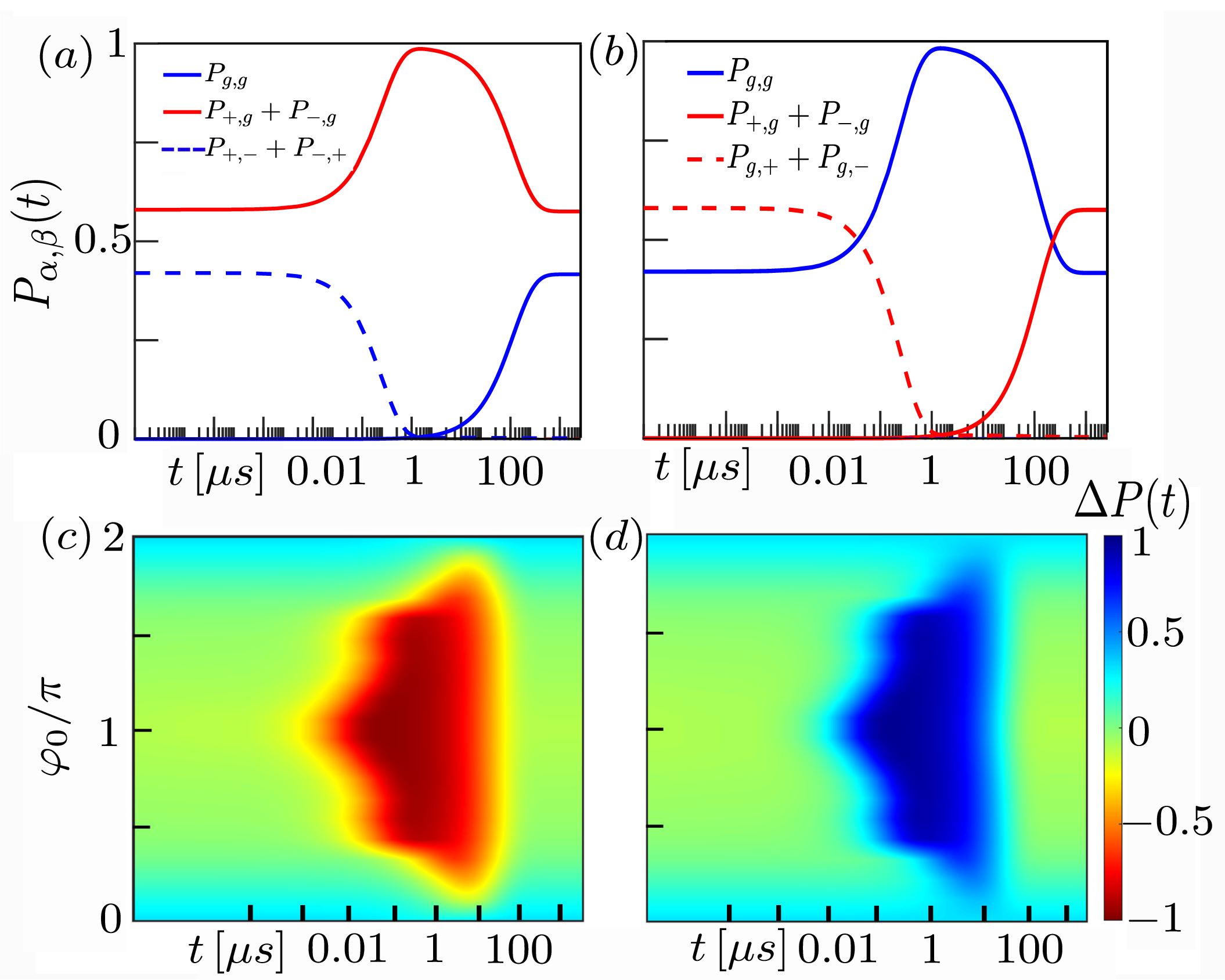}
\caption{Dynamical polarization after the initial microwave pulse, see Fig.~\ref{fig1}, for the parameters in Fig.~\ref{fig2} with $T_{\rm qp}=0.15\Delta$ and $T_{\rm env}=0.07\Delta$. 
The shown results follow from the simplified approach in Sec.~\ref{sec3a}.
(a) Dominant occupation probabilities $P_{\alpha,\beta}(t)$ vs $t$ (time in $\mu$s, on logarithmic scale) for $\varphi_0=\pi$ and the case shown in Fig.~\ref{fig1}(c), resulting in
dynamical polarization of the odd-parity sector. Even-parity (odd-parity) states are shown in blue (red) color.  Dashed curves indicate excited states in the respective sector.  
(b) Same but for polarization of the even-parity sector, see Fig.~\ref{fig1}(d). 
(c) Color-scale plot of $\Delta P(t)=P_{\rm even}(t)-P_{\rm odd}(t)$ in the 
$t$-$\varphi_0$ plane for the case in panel (a). 
(d) Same as in panel (c) but for polarization of the even-parity sector.
}
\label{fig3}
\end{figure}

\section{Dynamical parity polarization}\label{sec3}

In this section, using the theory framework in Sec.~\ref{sec2}, we provide a detailed discussion of dynamical parity polarization effects
after an initial microwave driving pulse whose frequency $\Omega_d$ matches a specific (even or odd) transition. 
In Sec.~\ref{sec3a}, we introduce and analyze a simplified picture for the time evolution of the state populations.   
We argue in Sec.~\ref{sec3b} that in practice, resonant driving of a single transition is possible despite of spurious state degeneracies
of the ABS levels predicted by our model.  
Subsequently, in Sec.~\ref{sec3c}, we consider a more realistic description of the driving pulse. However, we find qualitatively similar results
as from the simplified picture.

\subsection{Simplified picture for the pulse} \label{sec3a}

We start by imagining that the lower state is emptied by the resonant pulse such that the corresponding steady-state population is completely transferred to the initial excited-state population. Assuming complete population inversion in the respective parity sector while leaving the other one affected, the initial state ${\bf P}(0)$ immediately after the pulse  preserves the overall steady-state populations, $P_{\rm even}(0)=P_{\rm even}(\infty)$ and likewise for $P_{\rm odd}$.  We then arrive at initial conditions for the
matrix rate equation \eqref{dotP}
which simulate the effects of a resonant driving pulse, see~Eqs.~\eqref{initcond1} and \eqref{initcond}.

In  Fig.~\ref{fig3} we illustrate the resulting time evolution after the driving pulse. 
Panels (a) and (c) correspond to the case where the driving pulse matches the mixed-pair transition in the even parity sector (at fixed $\varphi_0$) and the odd-parity stabilization mechanism in Fig.~\ref{fig1}(c) operates: due to the energetic proximity to the continuum, the relaxation from the excited state $|\sigma,\bar \sigma\rangle$ proceeds much faster to the odd-parity ground state $|\sigma, g\rangle$ than to the even-parity ground state $|g,g\rangle$ since the condition 
\begin{equation}
   \frac{\Gamma_{|\sigma\bar\sigma\rangle \rightarrow |\sigma,g\rangle}}{\Gamma_{|\sigma,\bar\sigma\rangle \rightarrow |g,g\rangle}} \sim \frac{|I_{\varepsilon_2,\Delta}|^2}{|I_{\varepsilon_2,-\varepsilon_1}|^2} 
   \frac{J(\Delta-\varepsilon_2) e^{-(\Delta-\varepsilon_2)/T_{\rm env}}}{J(\varepsilon_2+\varepsilon_1)} \gg 1
\label{rates-ratio}
\end{equation}
is satisfied in this parameter range.  The validity of Eq.~\eqref{rates-ratio} is mainly due to the smallness of the 
current matrix element $I_{\varepsilon_2,-\varepsilon_1}$ for phase differences near $\varphi_0=\pi$. 
Formally, we find $I_{\varepsilon_2,-\epsilon_1}=0$ for $\varphi_0=\pi$ from Eq.~\eqref{calIAL} for the case of a symmetric junction,
which can be rationalized by noting that the involved ABS wave functions have opposite parity.
Although for our parameter choice, the spectral function $J(\varepsilon_2+\varepsilon_1)$ in the denominator of Eq.~\eqref{rates-ratio}
is dominated by the Ohmic contribution, its precise value is not essential for the dynamical polarization effect as long as it does not 
exceed the resonance peak in $J(\omega)$ at $\omega \sim \Omega$ which determines the $|\sigma,\bar\sigma\rangle\rightarrow|\sigma,g\rangle$ transition rate.
Similar estimates as in Eq.~\eqref{rates-ratio} also apply for the even-parity stabilization mechanism in Fig.~\ref{fig1}(b).

While the dynamical polarization stabilization  is most effective for $\varphi_0\sim \pi$, it remains robust over a broad phase range as shown in panel (c). 
We note that side features with enhanced parity polarization are seen for $\Omega\approx \Delta-\varepsilon_2(\varphi_0)$. 
For the chosen parameter set, we obtain dynamical odd-parity polarizations $P_{\rm odd}\agt 0.98$ on
time scales of order 100$\mu$s. The Ohmic background in $J(\omega)$, although not essential for the main effect, 
is necessary to obtain time scales as in the experiments of Ref.~\cite{Wesdorp2022}.
Similarly, panels (b) and (d) illustrate the corresponding effect 
when exciting the odd-parity transition $|\sigma ,g\rangle \rightarrow |g,\sigma\rangle$. 
In this case one obtains almost full dynamical polarization in the even-parity sector, which persists on a similar time scale $\sim 100\mu$s. 
We note that the above dynamical polarization mechanism remains robust over a 
broad temperature range. However, particularly strong polarization effects are obtained 
for $T_{\rm qp}\agt T_{\rm env}$, where only the ground states in each parity sector are 
populated in the steady state.  The couplings $\kappa$ and $\alpha_0$ then control the parity lifetimes.
   
\begin{figure*}
\includegraphics[scale=0.35]{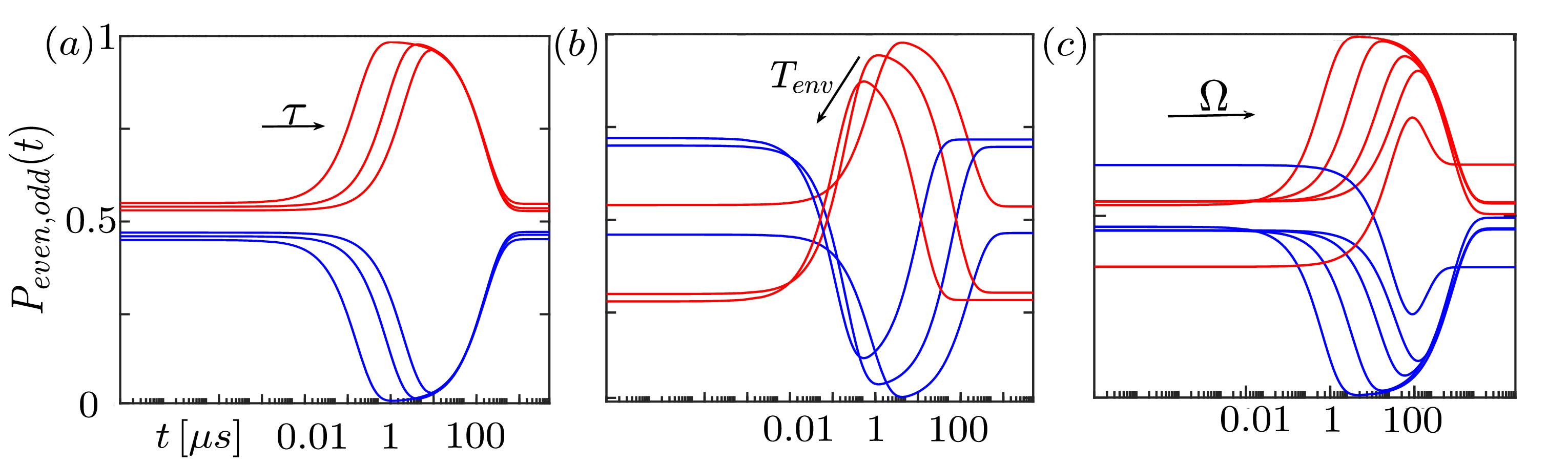}
\caption{Evolution of $P_{\rm odd}$ (red) and $P_{\rm even}$ (blue) vs time (in $\mu$s and on a logarithmic scale) after an initial 
microwave pulse inducing the mixed even transition, $|g,g\rangle \rightarrow |\sigma,\bar{\sigma}\rangle$. 
We use the parameters in Eq.~\eqref{parameters} together with $\ell=1.3$, $\varphi_0=\pi$, and $T_{\rm qp}=0.15\Delta$. 
The shown results follow from the simplified approach in Sec.~\ref{sec3a}. 
Panel (a) shows results for different transmission probabilities, ${\cal T}=0.45, 0.7, 0.9$, with $T_{\rm env}=0.07\Delta$.
Panel (b) illustrates the temperature dependence, $T_{\rm env}/\Delta=0.07, 0.15, 0.22$, with ${\cal T}=0.76$. 
Panel (c) shows results for different resonator frequencies, $\Omega/\Delta=0.03, 0.2, 0.3, 0.7, 0.99$, with $T_{\rm env}=0.07\Delta$ and ${\cal T}=0.76$.  
Arrows indicate increasing values for ${\cal T}$, $T_{\rm env}$, and $\Omega$, respectively. }  
\label{fig4}
\end{figure*}

In order to test the robustness of the dynamical polarization effect with respect to changes in various parameters,  
Fig.~\ref{fig4} shows the time evolution of $P_{\rm even, odd}(t)$ after a driving pulse inducing the $|g,g\rangle \rightarrow |\sigma,\bar{\sigma}\rangle$ transition. Panel (a) studies the effect of changing 
the transmission probability ${\cal T}$ of the weak link. We observe that the achievable dynamical polarization is rather insensitive to ${\cal T}$, 
provided that the condition $\Omega > \Delta - \varepsilon_2(\varphi_0)$ is met. 
However, with increasing ${\cal T}$, the time span during which the polarization of the odd-parity sector persists
becomes shorter. Panel (b) shows the evolution of $P_{\rm even/odd}(t)$ with increasing environmental temperature $T_{\rm env}$,
where the growing thermal imbalance between the steady-state populations with even and odd parity
comes along with a gradual decrease of the dynamical polarization. 
In particular, for $T_{\rm env}>T_{\rm qp}$, the state $|e,g\rangle$  has increasingly 
large population probability. Finally, in panel (c), we monitor $P_{\rm even/odd}(t)$ as 
the resonator frequency $\Omega$ is varied, where we find a decrease in the overall polarization 
with increasing $\Omega$.  One can rationalize this observation by noting that at fixed temperature 
$T_{\rm env}$, the average photon number in the resonator will decrease if $\Omega$ is increased. 
Since fewer resonator photons are available for promoting a trapped quasiparticle into the continuum, the dynamical polarization will be reduced. In principle, one could increase it again by simply
raising $T_{\rm env}$.  However, as detailed above, one needs to satisfy 
$T_{\rm qp}\agt T_{\rm env}$ at the same time.

\begin{figure*}
\includegraphics[scale=0.375]{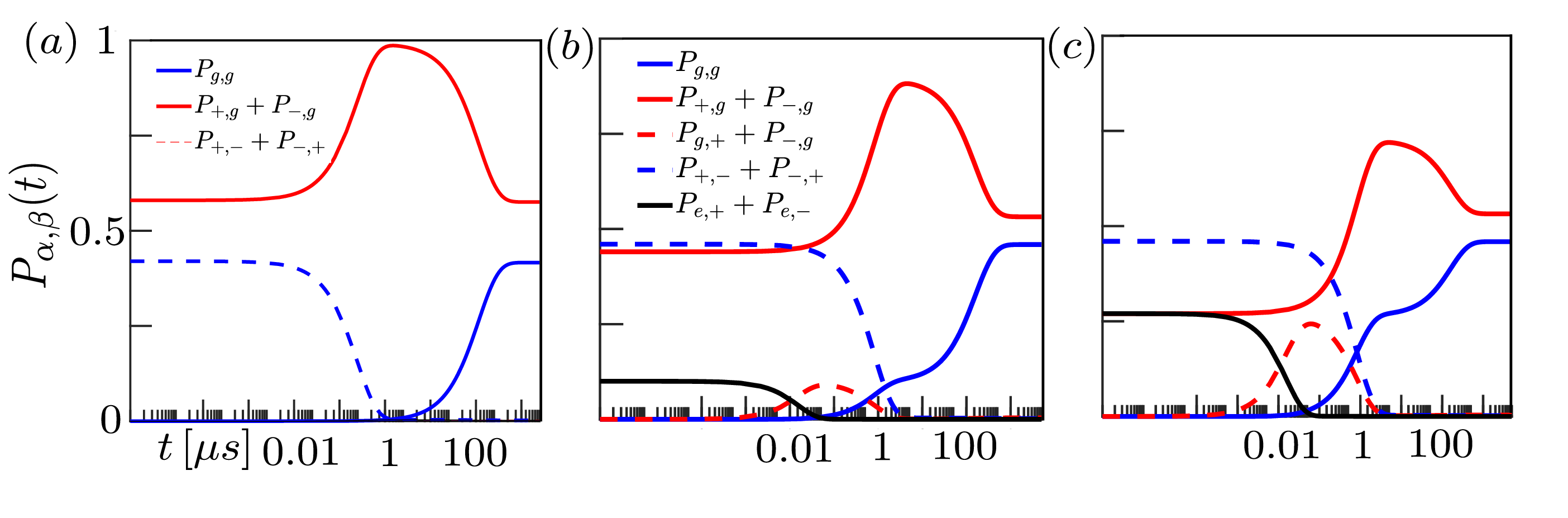}
\caption{Dynamical polarization after the initial microwave pulse for the same parameters as in Fig.~\ref{fig3}(a), again using the simplified approach in
Sec.~\ref{sec3a}.
Panel (a) reproduces Fig.~\ref{fig3}(a). Panels (b) and (c) illustrate the detrimental effect of
simultaneously driving the transitions $|g,g\rangle \leftrightarrow |\sigma,\bar{\sigma}\rangle$ and  $|\sigma,g\rangle \leftrightarrow |e,\sigma\rangle$, 
which are degenerate in the absence of interactions. 
We use the initial condition \eqref{initcond} with (b) $\tilde{P}_{e,\sigma}(0)=0.1$ and $\tilde{P}_{\sigma,g}(0)=0.44$, or (c) 
$\tilde{P}_{e,\sigma}(0)=\tilde{P}_{\sigma,g}(\infty)/2$ and $\tilde{P}_{\sigma,g}(0)=\tilde{P}_{\sigma,g}(\infty)/2$, see App.~\ref{appB}.} 
\label{fig5}
\end{figure*}

\begin{figure*}
\includegraphics[scale=0.35]{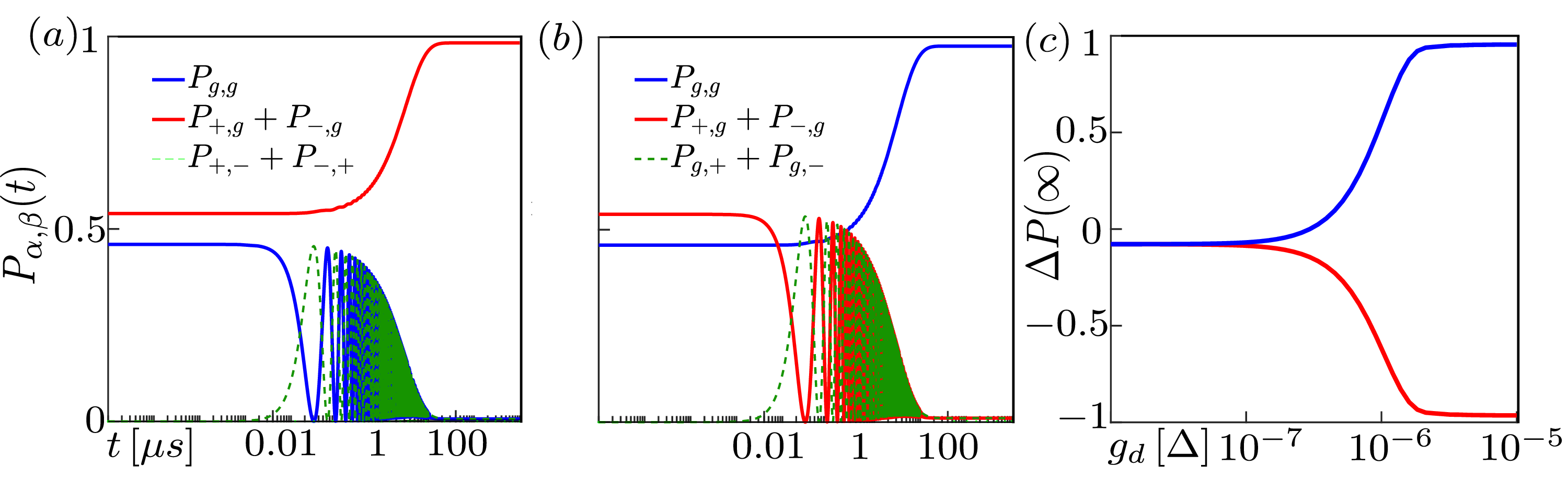}
\caption{Population dynamics, $P_{\alpha,\beta}(t)$, under a 
constant drive of amplitude $g_d=0.005\Delta$ for the parameters in Fig.~\ref{fig3}(a), obtained from
the approach in Sec.~\ref{sec3c}.
Panel (a) assumes a frequency $\Omega_d$ in resonance with 
the transition $|g,g\rangle \to |\sigma,\bar{\sigma}\rangle$. 
Panel (b) assumes $\Omega_d$ such that the transition $|\sigma,g\rangle \to |g,\sigma\rangle$ is driven. 
Panel (c) shows the dependence of $\Delta P(\infty)=P_{\rm even}(\infty)-P_{\rm odd}(\infty)$ on the drive amplitude 
$g_d$.  The red [blue] line is for $\Omega_d$ chosen as in panel (a) [(b)]. } 
\label{fig6}
\end{figure*}

\begin{figure*}
\includegraphics[scale=0.4]{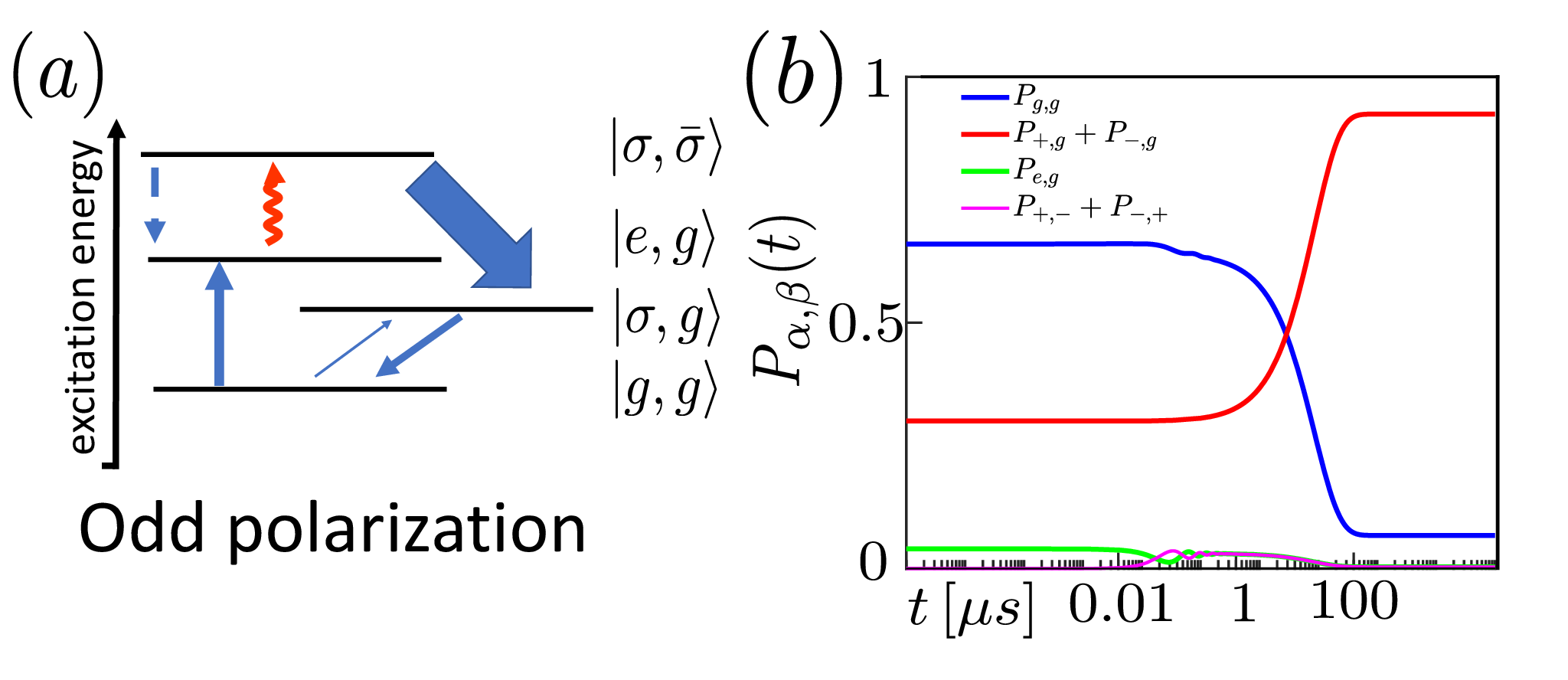}
\caption{Population dynamics under a constant drive of the $|e,g\rangle \to |\sigma,\bar \sigma\rangle$ transition, obtained from the approach in Sec.~\ref{sec3c}.
Panel (a) illustrates the relevant transition rates between many-body Andreev states, cf.~Fig.~\ref{fig1}. The driving-induced rate is shown as red arrow.
Panel (b) shows $P_{\alpha,\beta}(t)$ for $g_d=0.005\Delta$, using the parameters in Fig.~\ref{fig3}(a) and $T_{\rm env}=0.15\Delta$. } 
\label{fig7}
\end{figure*}

\subsection{Spurious state degeneracies}\label{sec3b}

The ABS dispersion relation found from our model, cf.~Eq.~\eqref{ALdispeq}, 
predicts certain degeneracies between even and odd sector transitions. 
For instance, the $|g,g\rangle \leftrightarrow |\sigma, \bar{\sigma}\rangle$ transition in the even sector is 
degenerate with the $|\sigma,g\rangle \leftrightarrow |e,\sigma\rangle$ transition in the odd sector.
Similarly, the $|\sigma,g\rangle \leftrightarrow |g,\sigma\rangle$ transition in the odd sector 
matches the $|e,g\rangle \leftrightarrow |\sigma,\bar{\sigma}\rangle$ transition in the even sector. 
If the drive frequency $\Omega_d$ of the microwave pulse corresponds to degenerate transitions, 
the initial conditions for the matrix master equation, see Eqs.~\eqref{initcond1} and \eqref{initcond}, 
have to be adapted accordingly.

In practice, however, such degeneracies are expected to be 
removed by spin-orbit coupling and Coulomb interaction effects.
Indeed, according to Ref.~\cite{Matute2022}, even a very weak Coulomb interaction,
as expected for the open junction regime studied in Refs.~\cite{Matute2022,Wesdorp2022},
will remove degeneracies in the even-sector transitions and change 
the relative position of the transition lines in the even and odd sectors. 
Taking as a reference the charging energy values quoted in Ref.~\cite{Hays2020}, 
we estimate that the $|g,g\rangle\to|\sigma,\bar\sigma\rangle$ and the $|\sigma,g\rangle
\to |e,\sigma\rangle$ transitions exhibit an energy splitting of the order of a few $\mu$eV,
due to the Coulomb energy cost penalizing a double occupation of the 
lower ABS manifold in the $|e,\sigma\rangle$ state. Although giving 
precise expressions for this splitting as a function of the model parameters is difficult, 
we safely estimate it to exceed the typical energy resolution in the present experiments. 
Leaving aside fine-tuned values of $\varphi_0$ where accidental degeneracies occur, we conclude that
the simultaneous driving of degenerate transitions is not expected in actual experiments.
For the results shown in this work, we have assumed that degeneracies are lifted by such a mechanism.  

Nonetheless, Fig.~\ref{fig5} illustrates the detrimental effects that such simultaneous excitations 
can have on the dynamical polarization effect if a degeneracy is actually present.
The extension of our theory to include spin-orbit coupling and/or Coulomb interaction effects 
is an important task for future research. 

\subsection{More realistic model for the driving pulse}\label{sec3c}

So far we have assumed a simplified picture where the microwave driving pulse is simulated by means of appropriate initial conditions for Eq.~\eqref{dotP}.
However, in actual experiments, see, e.g., Ref.~\cite{Wesdorp2022}, driving pulse effects could of course be more complex,
and one may need to account for a finite pulse duration, the precise drive amplitude, or the detailed pulse shape.
In the following, as a first step beyond the idealized approach in Sec.~\ref{sec3a}, 
we study the ABS population dynamics under a continuous and resonant microwave drive 
with drive amplitude $g_d$, where the drive frequency $\Omega_d$ matches a transition between two specific 
many-body Andreev states $|\alpha_1,\beta_1\rangle$ and $|\alpha_2,\beta_2\rangle$ belonging to the same parity sector,
\begin{equation}\label{resonant}
\Omega_d=E_{\alpha_2,\beta_2}-E_{\alpha_1,\beta_1},
\end{equation}
with the respective energies $E_{\alpha_1,\beta_1}<E_{\alpha_2,\beta_2}$.
We here assume that the drive does not couple to other transitions. In particular, degeneracies should be lifted as described in Sec.~\ref{sec3b}.
For $\Omega_d>\Delta-\varepsilon_2$, the driving field could in principle also cause transitions between the Andreev sector and the quasiparticle continuum.  
However, unless the corresponding drive-induced transition rates 
exceed the fast (drive-independent) parity-changing rates shown in Fig.~\ref{fig1}(c,d), such transitions will not have a qualitative impact on our results.
In what follows, we consider  weak drive amplitudes, $g_d\ll \Delta$, where this assumption is justified.

In order to model the continuous drive, we replace the effective Hamiltonian 
in the Lindblad equation \eqref{Lindbleq3} for the Andreev sector, $H_0\to H=H_0+H_d$. In the rotating wave approximation (RWA) and using the interaction picture, the drive Hamiltonian $H_d$ for the frequency $\Omega_d$ in Eq.~\eqref{resonant} is given by
\begin{equation}\label{drive}
H_d = g_d |\alpha_2,\beta_2\rangle\langle\alpha_1,\beta_1|+{\rm h.c.},
\end{equation}
where RWA is justified for $g_d\ll \Omega_d$.
As explained above, $H_d$ can then be truncated to include only transitions 
between the two many-body ABSs determining Eq.~\eqref{resonant}.
After projecting Eq.~\eqref{Lindbleq3} to the many-body ABS states $|\alpha,\beta\rangle$, 
we take into account only off-diagonal matrix elements of $\rho_A$ (``coherences'') between the two states connected
by the drive, $C_A^{|\alpha_1,\beta_1\rangle\to|\alpha_2,\beta_2\rangle}(t)$.
For the three transitions studied below, these complex-valued quantities are defined as 
\begin{eqnarray}  
C_{A}^{|g,g\rangle\to |\sigma,\bar{\sigma}\rangle}&=&\frac12 \left(\langle g,g|\rho_A|\sigma,\bar{\sigma}\rangle +\langle g,g|\rho_A|\bar{\sigma},\sigma\rangle \right),\nonumber \\
C_{A}^{|\sigma,g\rangle\to |g,\sigma\rangle}&=&\frac12\left(\langle +,g|\rho_A|g,+\rangle +\langle -,g|\rho_A|g,-\rangle\right ),\nonumber \\
C_{A}^{|e,g\rangle\to |\sigma,\bar{\sigma}\rangle}&=&\frac12\left(\langle e,g|\rho_A|\sigma,\bar{\sigma}\rangle +\langle e,g|\rho_A|\bar{\sigma},\sigma\rangle\right ).
\end{eqnarray}
We next define a vector ${\bf W}(t)= ({\bf P}(t), C_A^{}(t), C_A^\ast(t) )^T$, which contains the occupation probabilities of the many-body ABSs
in ${\bf P}(t)$ as before but also includes the coherences $C_A(t)$ and $C_A^\ast(t)$ in the last two entries. 
Enlarging the matrix ${\bf M}$ in Eq.~\eqref{dotP} by zero columns and rows for the last two entries, one then finds that Eq.~\eqref{dotP} has to be replaced by  
\begin{equation}\label{dotW}
  {\bf \dot{W}}(t) =  ( {\bf M} + {\bf N}_d ) {\bf W}(t),   
\end{equation}
where the matrix ${\bf N}_d$ has nonzero entries only for the four columns and rows referring to the ABS populations 
$P_{\alpha_1,\beta_1}$ and $P_{\alpha_2,\beta_2}$ or to coherences. Explicit expressions are given in App.~\ref{appB}.

Numerical results for the ABS population dynamics under a continuous microwave drive of amplitude $g_d=0.005\Delta$ 
are shown in Fig.~\ref{fig6}(a,b).  In panel (a), the mixed even transition described above is resonantly driven, resulting
in a polarization of the odd sector.  We observe that the drive induces Rabi oscillations between the driven ABS states, which decay on a short time scale before the dynamical 
polarization behavior sets in.  Since we have a continuous drive, the polarization of the odd sector then persists for long times.
A similar picture is observed in panel (b), where $\Omega_d$ matches the mixed odd transition $|\sigma,g\rangle \to |g,\sigma\rangle$ 
and the even sector becomes polarized after an initial transient characterized by damped Rabi oscillations. For both cases,
the dependence of the long-time polarization, $\Delta P(\infty)=P_{\rm even}(\infty)-P_{\rm odd}(\infty)$,
on the drive amplitude is shown in Fig.~\ref{fig6}(c). We observe that a rather weak drive amplitude, $g_d\agt 10^{-6}\Delta$, is sufficient
to induce the polarization effect, and the phenomenon then remains insensitive to the precise value of $g_d$.

We therefore find basically the same behavior as obtained from the idealized initialization procedure in Sec.~\ref{sec3a}.  
In particular, the respective peak polarization values are very close to the steady-state polarization values found under a continuous resonant 
drive with $10^{-6}\Delta <g_d\ll\Omega_d$.  It is worth emphasizing that
Fig.~\ref{fig6} also shows that our idealized initialization approach implicitly requires a sufficiently long pulse duration compared to the decay time of the Rabi oscillations.  
(We note in passing that the latter time is determined by the inverse of the rate $\tilde \Gamma_A$ in Eq.~\eqref{gammaA}.)
The above results are consistent with key observations made in Ref.~\cite{Wesdorp2022}, i.e., 
 optimal polarization requires a certain minimal drive amplitude and decreases with the delay time. 
 We cannot exclude, however, that multi-photon processes (not included in our modeling) are needed for a more quantitative explanation
 of the results of Ref.~\cite{Wesdorp2022}.  In particular, the pumping of resonator photons due to a strong drive power is not accounted for
 by our theory.  This effect can result in multiple photon replicas of
 transition lines and has presumably been observed in Ref.~\cite{Wesdorp2022}.

Finally, we consider the effects of a continuous driving of transitions between higher-energy states.
We have discussed above how to polarize the odd sector by driving the $|g,g\rangle \to |\sigma,\bar{\sigma}\rangle$ transition.
However, as illustrated in Fig.~\ref{fig7} for an elevated temperature $T_{\rm env}$, if a small but finite steady-state population of the state $|e,g\rangle$ is present, a polarization of the odd-parity sector 
can alternatively be generated by the smaller drive frequency, $\Omega_d\approx \varepsilon_2-\varepsilon_1$.  (For the case in Fig.~\ref{fig3}(a), this also implies $\Omega_d\approx \Delta-\varepsilon_1$.)
While in the absence of interactions, this transition is degenerate with the $|\sigma,g\rangle \to |g,\sigma\rangle$ transition in the odd sector, 
even mild interactions can split both transitions.  Such a mechanism may account for the behavior observed in Ref.~\cite{Wesdorp2022} when driving with frequencies $\sim 29$~GHz, 
just above the odd parity transition lines. 
We note that at elevated temperatures, we find that a polarization in the odd sector can also be 
generated by driving the excited transition $|e,g\rangle\to |\sigma,\bar{\sigma}\rangle$, with drive frequency $\Omega_d=\varepsilon_2-\varepsilon_1$, which may help to explain experimental observations in Ref.~\cite{Wesdorp2022}.

\section{Conclusions}\label{conc}

Our theory suggests a simple physical mechanism for inducing almost perfect dynamical parity polarization of the subgap states in finite-length Josephson junctions. 
This effect should be useful in facilitating Andreev qubit quantum manipulations within a protected parity subspace.
Although we have considered a simplified model which neglects many aspects of realistic devices, 
the mechanism discussed here provides a basis for qualitatively understanding the intriguing experimental 
results of Ref.~\cite{Wesdorp2022}. 
Our predictions on the phase dependence, see Figs.~\ref{fig3}(c,d), as well as on the 
dependence on temperature $T_{\rm env}$ and resonator frequency $\Omega$, 
could be tested by future experiments, where the proposed mechanism also 
offers a simple strategy for optimizing the parity polarization.  
Our formalism directly allows to consider superconducting weak links harboring more than two ABSs as
well as spin-orbit coupling effects \cite{Tosi2019,Park2017}. 
Future work should clarify the role of Coulomb interactions, e.g., by modeling the weak link as a quantum dot \cite{Alvaro2011,Kurilovich2021,Hermansen2022}.

\begin{acknowledgments}
We wish to thank J. Cerillo, M.~Goffman, B.~van Heck and J.~J.~Wesdorp for discussions. 
We acknowledge funding by FET-Open contract AndQC, the Deutsche Forschungsgemeinschaft (DFG, German Research Foundation) Grant No.~277101999 - TRR 183 (project C01), Grant No.~EG 96/13-1, and under Germany's Excellence Strategy - Cluster of Excellence Matter and Light for Quantum Computing (ML4Q) EXC 2004/1 - 390534769, and by the Spanish AEI through Grant No.~PID2020-117671GB-I00 and through the ``Mar\'{\i}a de Maeztu'' Programme for Units of Excellence in R\&D (CEX2018-000805-M).
\end{acknowledgments}

\appendix
\section{BdG eigenstates} \label{appA}

We consider a  single-channel superconducting weak link described by the Hamiltonian (\ref{HLR})
together with a transfer matrix 
\begin{eqnarray}\nonumber
\hat T &=& \frac{1}{\sqrt{\cal T}} \left( \begin{array}{cc} e^{-i \tau_z \eta \hat \theta_R} & 0  \\ 0 & e^{i \tau_z \eta \hat \theta_R} \end{array} \right)\left( \begin{array}{cc} 1 & r  \\ r & 1 \end{array} \right) \times \\
\label{BC2} &&\qquad \times
\left( \begin{array}{cc} e^{-i \tau_z (1-\eta) \hat \theta_R} & 0  \\ 0 & e^{i \tau_z (1-\eta) \hat \theta_R} \end{array} \right),
\end{eqnarray}
which connects the envelope functions on both sides, $\Psi(0^-,t)=\hat T\Psi(0^+,t)$.
The parameter $\eta \in [0, 1]$ determines the position of a local elastic scatterer in the normal-conducting weak link. We here consider the symmetric case $\eta = 1/2$, where Eq.~\eqref{BC2} reduces to Eq.~\eqref{BC}. However,
it is straightforward to adapt our formalism for $\eta\ne 1/2$.   

With the gauge choice $\phi_{j}(t) = s_j \varphi(t) /2$, where $s_{j=L/R}=\pm$, and writing $\varphi(t)=\varphi_0+\tilde\varphi(t)$, we next expand $H_{L/R}$ in Eq.~\eqref{HLR} to linear order in the small phase fluctuations $\tilde \varphi(t)$ caused by the electromagnetic environment.
Applying the Josephson relation, we can use the auxiliary relation 
\begin{eqnarray}
 && V_j \tau_z + \Delta \tau_x e^{i \tau_z \phi_j} = e^{-is_j\tau_z\varphi_0/4} \times \\  \nonumber && \times  \left(
\frac{s_j \dot {\tilde \varphi}}{ 4} \tau_z + \frac{s_j\tilde \varphi \Delta}{2} \tau_y 
+\Delta \tau_x \right) e^{i s_j\tau_z \varphi_0/4} + {\cal O}\left( \tilde\varphi^2\right).
\end{eqnarray}
In the next step, we gauge away $\dot{\tilde\varphi}$ from the matching condition (where $\alpha=\pm$),
\begin{equation}
    \psi_{\alpha}(0^\pm,t)\to e^{ \pm i \alpha\tau_z \chi(t)/2} \psi_\alpha(0^\pm,t),
    \quad \chi=\frac{L}{4v_F}\dot{\tilde\varphi},
\end{equation}
taking into account that $e^{i\hat\theta_R(t)}\approx e^{i\hat \theta(t)} e^{i\chi(t)}\approx e^{i\chi(t)}e^{i\hat \theta(t)}$ and neglecting all derivatives $\partial^n_t \tilde\varphi$ with $n\ge 2$.  At this point, we can expand $H_j$ also to lowest order in $\dot{\tilde \varphi}$, and finally remove $\varphi_0$ from 
$H_{L/R}$ by a gauge transformation, which correspondingly
modifies the transfer matrix. 
After the above sequence of steps, we obtain $H(t) = H_{0} + H_c(t)$,
with the fermion Hamiltonian
\begin{equation}
H_{ 0} = \sum_{\pm} \int_{x \neq 0} dx \,  \psi^\dagger_\pm (x) \left(
\mp i  v_F \tau_z \partial_x + \Delta \tau_x \right) \psi^{}_\pm(x).
\end{equation}
This noninteracting problem is diagonalized by solving the BdG 
equations \eqref{BdG}. The eigenenergies $\varepsilon_\nu$ and the corresponding spinor eigenstates 
$\Psi_\nu(x)$ are specified below.  
We note that although the matching condition \eqref{BC} is nonlocal in time,
after the above expansion, we obtain a standard stationary matching condition, see Eq.~\eqref{BdG}.
We find from the above steps that the coupling between fermionic quasiparticles  and the 
electromagnetic environment is described by
\begin{eqnarray}\label{H1}
H_c(t) &=& \sum_{\pm} \int_{x \neq 0} dx \,  \psi^\dagger_\pm(x) U^{}_\pm(x, t) \psi^{}_\pm(x),\\
\nonumber
U^{}_\pm(x, t) &=& {\rm sgn}(-x)\left( \frac{\dot{\tilde \varphi}}{4} \, (\tau_z\mp \ell \tau_y) + \frac{ \tilde \varphi }{ 2} \Delta \tau_y \right),
\end{eqnarray}
where $\ell=L/\xi_0$. Using the Josephson current operator 
${\cal I} = \sum_{\nu,\nu'} {\cal I}_{\nu\nu'} \gamma_{\nu}^\dagger\gamma_{\nu'}^{}$ with the matrix elements \eqref{calI}, we   arrive at
\begin{equation}
H_c(t)  = \frac{\tilde \varphi(t)}{2}  {\cal I}(t) + \partial_t (\ldots ),
\label{H1calI}
\end{equation}
where the time derivative term does not affect the dynamics and can be omitted.

We next specify the quasiparticle spinor wave functions, $\Psi_\nu(x)$, and the corresponding eigenenergies, $\epsilon_\nu$, which follow by solving the BdG problem \eqref{BdG}. 
Here, $\Psi=(\psi_+,\psi_-)^T$ is a bispinor in Nambu and  left-right mover space.
Because of the matching condition at $x=0^\pm$, we have the unconventional normalization condition
\begin{eqnarray}\label{norm1}
   && \int_{-\infty}^\infty dx |\Psi_\nu(x)|^2 = 1-\zeta_\nu ,\\ \nonumber 
&&    \zeta_\nu = \frac{L}{2} \left( |\Psi_\nu(0^-)|^2 + |\Psi_{\nu}(0^+)|^2 \right),
\end{eqnarray}
with $L=\ell \xi_0$.  For continuum states, we find $\zeta_\nu\sim {\cal O}(L/L_{\rm sc})\to 0$, 
with the length $L_{\rm sc}\to \infty$ of the superconducting leads.
However, for ABS solutions, $\zeta_\nu$ must be accounted for in the normalization. In particular, 
$\zeta_\nu$ can depend on $\varphi_0$. For notational ease, we set $v_F = \Delta = 1$ below.

\emph{Andreev bound states.---}For $|\varepsilon| < 1$, we have subgap ABS solutions with $\nu\equiv \lambda$, which can be written as
\begin{equation}
\Psi_\lambda(x) = \Theta(-x) e^{\kappa x} \left( \begin{array}{l} a \psi_h \\ b \psi_e \end{array} \right) +
\Theta(x) e^{-\kappa x} \left( \begin{array}{l} c \psi_e \\ d \psi_h \end{array} \right),
\label{PsiE}
\end{equation}
where $\Theta(x)$ is the Heaviside step function and $\psi_{e/h}$ are 
electron/hole-type Nambu spinors for localized states satisfying
\begin{equation}\label{locspin}
\left( \varepsilon \mp i\kappa \tau_z -  \tau_x \right) \psi_{e/h} = 0.
\end{equation}
With $\gamma = \cos^{-1}\varepsilon \in (0, \pi)$, one finds
\begin{equation}\label{gamma}
\kappa = \sin \gamma = \sqrt{1 - \varepsilon^2},\quad \psi_{e/h} = \frac{e^{\pm i \tau_z \gamma/2}}{ \sqrt{2}}\left( \begin{array}{l} 1 \\ 1 \end{array} \right).
\end{equation}
From Eq.~\eqref{norm1}, the scalar amplitudes ($a,b,c,d)$ in Eq.~\eqref{PsiE} obey the normalization condition 
$|a|^2 + |b|^2 + |c|^2 + |d|^2 = 2 \kappa/(1 +\ell \kappa)$.
The matching equations in Eq.~\eqref{BdG} then yield the relations 
\begin{eqnarray}\nonumber
\sqrt{\cal T} \, a & = & e^{i \tau_z \gamma_+} c + r e^{i \tau_z \varphi_0 / 2} d,  \\
\sqrt{\cal T} \, b & = & r e^{i \tau_z \varphi_0 / 2} c + e^{-i \tau_z \gamma_-}  d,
\end{eqnarray}
where $\gamma_\pm = \gamma - \theta \pm \varphi_0 / 2$ and $\theta=\ell\varepsilon$. Hence
$c$ and $d$ follow from
\begin{equation}
\left( \begin{array}{cc} \sin \gamma_+ & r \sin (\varphi_0/2) \\
r \sin (\varphi_0/2) & - \sin \gamma_- \end{array} \right) \left( \begin{array}{l} c \\ d \end{array} \right) = 0.
\end{equation}
As a consequence, the dispersion equation for the phase-dependent ABS levels, 
$\varepsilon=\varepsilon_\lambda(\varphi_0)$, takes the form in Eq.~\eqref{ALdispeq} 
with $m\in \mathbb{Z}$.
  For each solution $\varepsilon_\lambda$ 
with $m=m_\lambda$, another root of Eq.~\eqref{ALdispeq} is given by $-\varepsilon_\lambda$ with 
$m=1-m_\lambda$.

\emph{Above-gap continuum states.---}We next turn to above-gap states 
with quantum numbers $\nu\equiv p=(\varepsilon,s)$ and $|\varepsilon| > 1$. Scattering state solutions for continuum quasiparticles can be written as the  
sum of incoming and outgoing plane waves,
$\Psi_{p}(x) = \Psi^{(\rm in)}_{p}(x) + \Psi^{(\rm out)}_{p}(x)$.
There are four incoming states labeled by $s \in \{ 1,2,3,4 \}$,
which describe electron- or hole-type states injected from the left or right side,
\begin{eqnarray}\nonumber
&& \Psi^{(\rm in)}_{p}(x) = \Theta(-x) \frac{e^{i k x}}{\sqrt{L_{\rm sc}}}
\left[ \delta_{s,1} \left( \begin{array}{l} \tilde \psi_e \\ 0 \end{array} \right) +
\delta_{s,2} \left( \begin{array}{l} 0 \\ \tilde \psi_h \end{array} \right) \right]\\
&&\quad +\,
\Theta(x) \frac{e^{-i k x} }{ \sqrt{L_{\rm sc}}}
\left[ \delta_{s,3} \left( \begin{array}{l} \tilde \psi_h \\ 0 \end{array} \right) +
\delta_{s,4} \left( \begin{array}{l} 0 \\ \tilde \psi_e \end{array} \right) \right],
\label{Psiin}
\end{eqnarray}
and the respective four outgoing states,
\begin{equation}
\Psi^{(\rm out)}_{p}(x) = \Theta(-x) \frac{e^{-i k x}}{ \sqrt{L_{\rm sc}}}
\left( \begin{array}{l} a_p \tilde \psi_h \\ b_p \tilde \psi_e \end{array} \right) +
\Theta(x) \frac{e^{i k x} }{\sqrt{L_{\rm sc}}}
\left( \begin{array}{l} c_p \tilde \psi_e \\ d_p \tilde \psi_h \end{array} \right).
\label{Psiout}
\end{equation}
Here $\tilde \psi_{e/h}$ are electron/hole-type Nambu spinors for continuum states 
 satisfying Eq.~\eqref{locspin} with $i\kappa\to k$.
Introducing $\tilde \gamma = \sinh^{-1}(\sqrt{\varepsilon^2 - 1} )\in [0, \infty)$ 
and $\sigma_\varepsilon= {\rm sgn}(\varepsilon)$, we write
\begin{equation}\label{aux1}
\varepsilon = \sigma_\varepsilon\cosh \tilde \gamma, \quad k = \sigma_\varepsilon \sinh \tilde \gamma,\quad
\tilde \psi_{e/h} = \frac{e^{\pm \tau_z \tilde \gamma/2} }{ \sqrt{2 \cosh \tilde \gamma}} 
\left( \begin{array}{l} 1 \\ \sigma_\varepsilon \end{array} \right).
\end{equation}
Using
\begin{equation}
\sqrt{1 - (\varepsilon + i 0^+)^2} =   \kappa(\varepsilon) \Theta(1-|\varepsilon|) - i k(\varepsilon) \Theta(|\varepsilon| - 1 ),
\end{equation}
we observe that the eigenstates \eqref{PsiE} and \eqref{Psiout} are related by analytic 
continuation in the complex-energy plane, and the scalar amplitudes $(a_p, b_p, c_p, d_p)$ 
again follow from matching conditions.  With $\theta =\ell \varepsilon$, 
\begin{eqnarray}\nonumber
\sqrt{\cal T} \left( \begin{array}{l} a_p \tilde \psi_h + \delta_{s,1} \tilde \psi_e \\
b_p \tilde \psi_e + \delta_{s,2} \tilde \psi_h \end{array} \right) &=& e^{i \tau_z \varphi_0/2}
\left( \begin{array}{cc} e^{-i \tau_z \theta} & r  \\ r & e^{i \tau_z \theta} \end{array} \right)
\\ &\times& \left( \begin{array}{l} c_p \tilde \psi_e + \delta_{s,3} \tilde \psi_h \\
d_p \tilde \psi_h + \delta_{s,4} \tilde \psi_e \end{array} \right),
\label{bccont}
\end{eqnarray}
with the normalization $|a_p|^2 + |b_p|^2 + |c_p|^2 + |d_p|^2 = 1$.

\emph{Supercurrent matrix elements.---}We next provide explicit expressions for the
current matrix elements in Eq.~\eqref{calI}.
First, for two ABSs with energies $\varepsilon_\lambda$ and $\varepsilon_{\lambda'}$, using $\gamma_{\lambda}=\cos^{-1} \varepsilon_{\lambda}$ and similarly for $\gamma_{\lambda'}$, 
 we employ the relations 
\begin{eqnarray}\nonumber
\psi^{(\lambda)\, \dagger}_{e/h} \tau_z \psi^{(\lambda')}_{e/h} &=&
\mp i \, \sin \left( \gamma_\lambda - \gamma_{\lambda'} \over 2 \right),\\
\psi^{(\lambda)\, \dagger}_{e/h} \tau_y \psi^{(\lambda')}_{e/h} &=&
\mp \sin \left( \gamma_\lambda + \gamma_{\lambda'} \over 2 \right).
\end{eqnarray}
The matrix elements for ABS-ABS transitions follow with $\omega_{\lambda\lambda'}=\varepsilon_\lambda-\varepsilon_{\lambda'}$ in the form
\begin{widetext}
\begin{eqnarray}
{\cal I}_{\lambda \lambda'} & =&
\left[ \frac{\omega_{\lambda \lambda'}}{2} \, \sin \left( \gamma_\lambda - \gamma_{\lambda'} \over 2 \right) +
\sin \left( \gamma_\lambda + \gamma_{\lambda'} \over 2 \right)
\right]  \frac{
a^\ast_\lambda a_{\lambda'} - b^\ast_\lambda b_{\lambda'} + c^\ast_\lambda c_{\lambda'} - d^\ast_\lambda d_{\lambda'}}
 {\sin\gamma_\lambda + \sin\gamma_{\lambda'}} \nonumber
\\ \label{calIAL}
  &+ &i\ell\frac{\omega_{\lambda\lambda'}}{2}   \sin \left( \gamma_\lambda + \gamma_{\lambda'} \over 2 \right) \frac{a^\ast_\lambda a_{\lambda'} + b^\ast_\lambda b_{\lambda'} + c^\ast_\lambda c_{\lambda'} + d^\ast_\lambda d_{\lambda'}}{\sin\gamma_{\lambda} +
\sin\gamma_{\lambda'}} .
\end{eqnarray}
We next observe that in the limit $L_{\rm sc} \to \infty$, matrix elements between continuum states, $\Psi_p$ and $\Psi_{p'}$, can only be finite  if $\varepsilon_p =\varepsilon_{p'}$. 
However, the relations ${\tilde \psi}_e^\dagger \tau_y \tilde \psi_e = {\tilde \psi}_h^\dagger \tau_y \tilde \psi_h = 0$ imply that such matrix elements vanish as well. 
We conclude that phase fluctuations cannot induce transitions between continuum states, ${\cal I}_{p p'} = 0$.

We proceed with the matrix element connecting an ABS with energy $\varepsilon_\lambda$ to a continuum state with energy $\varepsilon_p$.  As in Eq.~\eqref{aux1}, we write $\varepsilon_p= \sigma_{\varepsilon_p}\cosh\tilde\gamma_p$ and $k_p=\sigma_{\varepsilon_p}\sinh\tilde\gamma_p$.  
The matrix elements follow by using the relations
\[
\psi_{e/h}^{(\lambda) \, \dagger } \tau_z \tilde \psi_{e/h} =
{w(\pm z^\ast) \over \sqrt{\cosh \tilde \gamma_p}},\quad
\psi_{e/h}^{(\lambda) \, \dagger} \tau_z \tilde \psi_{h/e} = {w(\mp z) \over \sqrt{\cosh \tilde \gamma_p}},\quad
\psi_{e/h}^{(\lambda)\, \dagger} \tau_y \tilde \psi_{e/h} = {i w(\pm z) \over \sqrt{\cosh \tilde \gamma_p}},\quad 
\psi_{e/h}^{(\lambda) \, \dagger} \tau_y \tilde \psi_{h/e} = {i w(\mp z^\ast) \over \sqrt{\cosh \tilde \gamma_p}},
\]
with 
\[
z = (\tilde \gamma_p+i\gamma_\lambda)/2, \quad \omega_{p\lambda}=\varepsilon_p-\varepsilon_\lambda,
\quad w(z) = \Theta(-\varepsilon_p) \cosh z + \Theta(\varepsilon_p) \sinh z,\quad 
W(z) = w(z) + \frac{\omega_{p\lambda}}{2} \, w^\ast(z).
\]
Note that $W(z^\ast) = W^\ast(z)$ and $W(-z) = - \sigma_{\varepsilon_p} W(z)$.
We then obtain 
\begin{eqnarray*}
{\cal I}_{\lambda p} &=& {i \over \sqrt{L_{\rm sc} \, \cosh \tilde \gamma_p}}
\left[ \frac{\left( a^\ast_\lambda a_p - d^\ast_\lambda d_p \right) W(-z) +
\left( b^\ast_\lambda b_p - c^\ast_\lambda c_p\right) W(z)} {\sin\gamma_\lambda - i k_p}  + \frac{\left( a^\ast_\lambda \delta_{s,1} - d^\ast_\lambda \delta_{s,4} \right) W(z^\ast) +
\left( b^\ast_\lambda \delta_{s,2} - c^\ast_\lambda \delta_{s,3} \right) W(-z^\ast) }
{ \sin\gamma_\lambda + i k_p} \right] \\
&+&\frac{\ell\omega_{p\lambda}}{2\sqrt{ L_{\rm sc} \, \cosh \tilde \gamma_p}}  \left[ \frac{\left( a^\ast_\lambda a_p + d^\ast_\lambda d_p \right) w(-z)  -
\left( b^\ast_\lambda b_p + c^\ast_\lambda c_p\right) w(z)}{\sin\gamma_\lambda - i k_p} +
\frac{\left( a^\ast_\lambda \delta_{s,1} + d^\ast_\lambda \delta_{s,4} \right) w(z^\ast) -
\left( b^\ast_\lambda \delta_{s,2} + c^\ast_\lambda \delta_{s,3} \right) w(-z^\ast)}{\sin\gamma_\lambda + i k_p} 
  \right].
\end{eqnarray*}
\end{widetext}
Finally, we note that summations over $p=(\varepsilon,s)$ are performed for $L_{\rm sc}\to \infty$ by using 
$\frac{1}{L_{\rm sc}} \sum_{p = (\varepsilon, s)} \cdots =
\int d \varepsilon \, \nu(\varepsilon) \sum_{s = 1}^4 \cdots$,
where $\nu(\varepsilon)$ is the BCS density of states (per unit length) in the leads,
$\nu(\varepsilon) = \frac{1}{2\pi} \frac{|\varepsilon|}{\sqrt{\varepsilon^2 - 1}} 
\Theta(|\varepsilon| - 1).$

\section{On matrix rate equations}\label{appB}

\begin{widetext}
\begin{table*}[t]
	\begin{tabular}{|l||c|c|c|c|c|c|c|c|c|c|c|}
		\hline
		& $gg$ & $\sigma g$ & $eg$ & $ge$ & $ee$ & $\sigma \sigma$ & $\sigma\bar\sigma$ &
		$g\sigma$ & $e \sigma$ & $\sigma e$ \\
		\hline \hline
		$gg$ & $-\tilde \Gamma_{gg}$ & $\Gamma^{\rm in}_{-1}$ &$\Gamma_{1,-1}$ & $\Gamma_{2,-2}$ & 0 & 0 &
		$\Gamma_{1,-2}$ & $\Gamma^{\rm in}_{-2}$ & 0 & 0 \\
		$\sigma g$ & $2\Gamma^{\rm in}_1$ & $-\tilde \Gamma_{\sigma g}$ & $2\Gamma^{\rm in}_{-1}$ & 0 & 0 & $\Gamma^{\rm in}_{-2}$ & $\Gamma^{\rm in}_{-2}$ &
		$\Gamma_{-1,-2}$ & $\Gamma_{1,-2}$ & $\Gamma_{2,-2}$ \\
		$eg$ & $\Gamma_{-1,1}$ & $\Gamma^{\rm in}_1$ & $-\tilde \Gamma_{eg}$ & 0 & $\Gamma_{2,-2}$ & 0 &
		$\Gamma_{-1,-2}$ & 0 & $\Gamma^{\rm in}_{-2}$ & 0 \\
		$ge$ & $\Gamma_{-2,2}$ & 0 & 0 & $-\tilde \Gamma_{ge}$ & $\Gamma_{1,-1}$ & 0 &
		$\Gamma_{1,2}$ & $\Gamma^{\rm in}_2$ & 0 & $\Gamma^{\rm in}_{-1}$ \\
		$ee$ & 0 & 0 & $\Gamma_{-2,2}$ & $\Gamma_{-1,1}$ & $-\tilde \Gamma_{ee}$ & 0 &
		$\Gamma_{-1,2}$ & 0 & $\Gamma^{\rm in}_2$ & $\Gamma^{\rm in}_1$ \\
		$\sigma \sigma$ & 0 & $\Gamma^{\rm in}_2$ & 0 & 0 & 0 & $-\tilde \Gamma_{\sigma \sigma}$ & 0 &
		$\Gamma^{\rm in}_1$ & $\Gamma^{\rm in}_{-1}$ & $\Gamma^{\rm in}_{-2}$ \\
		$\sigma\bar\sigma$ & $2\Gamma_{-1,2}$ & $\Gamma^{\rm in}_2$ & $2\Gamma_{1,2}$ & $2\Gamma_{-1,-2}$ & $2\Gamma_{1,-2}$ & 0 &
		$-\tilde \Gamma_{\sigma\bar\sigma}$ &
		$\Gamma^{\rm in}_1$ & $\Gamma^{\rm in}_{-1}$ & $\Gamma^{\rm in}_{-2}$ \\
		$g \sigma$ & $2\Gamma^{\rm in}_2$ & $\Gamma_{12}$ & 0 & $2\Gamma^{\rm in}_{-2}$ & 0 & $\Gamma^{\rm in}_{-1}$ & $\Gamma^{\rm in}_{-1}$ &
		$-\tilde \Gamma_{g \sigma}$ & $\Gamma_{1,-1}$ & $\Gamma_{1,-2}$ \\
		$e \sigma$ & 0 & $\Gamma_{-1,2}$ & $2\Gamma^{\rm in}_2$ & 0 & $2\Gamma^{\rm in}_{-2}$ & $\Gamma^{\rm in}_1$ & $\Gamma^{\rm in}_1$ &
		$\Gamma_{-1,1}$ & $-\tilde \Gamma_{e \sigma}$ & $\Gamma_{-1,-2}$\\
		$\sigma e$ & 0 & $\Gamma_{-2,2}$  & 0 & $2\Gamma^{\rm in}_1$ & $2\Gamma^{\rm in}_{-1}$ & $\Gamma^{\rm in}_2$ & $\Gamma^{\rm in}_2$ &
		$\Gamma_{-1,2}$ & $\Gamma_{1,2}$ & $-\tilde \Gamma_{\sigma e}$\\
		\hline
	\end{tabular}
	\caption{\label{tab2} Reduced $10\times 10$ rate matrix ${\bf M}$ appearing in Eq.~\eqref{dotP},
	expressed in terms of the transition rates $\Gamma_{\lambda\lambda'}$ with ABS indices 
	$\lambda,\lambda'\in\{\pm 1,\pm 2\}$, see Eq.~\eqref{Gammanumu}, and the rates
	$\Gamma^{\rm in/out}_\lambda$ in Eq.~\eqref{gammainout}. }
\end{table*}
\end{widetext}

We here discuss the matrix rate equation \eqref{dotP} in more detail.
In our calculations, we make use of ``spin'' degeneracies of the many-body Andreev levels
which allow us to reduce the effective matrix dimension for ${\bf M}$ down to ten.  (However, the resulting matrix is not symmetric anymore.) For this purpose, for $\alpha \in \{g,e\}$, we write
\begin{eqnarray}\label{red1}
    \tilde{P}_{\alpha,\sigma}&=&P_{\alpha,+}+P_{\alpha,-}, \quad \tilde{P}_{\sigma,\alpha}=P_{+,\alpha}+P_{-,\alpha},\\ \nonumber
    \tilde{P}_{\sigma,-\sigma}&=&P_{+,-}+P_{-,+},\quad \tilde{P}_{\sigma,\sigma}= P_{+,+}+P_{-,-}.  
\end{eqnarray} 
Defining the 10-component vector ${\bf P}(t)=
\left( P_{g,g}, \tilde{P}_{\sigma ,g}, P_{e,g}, P_{g,e}, P_{e,e},
\tilde{P}_{\sigma ,\sigma}, \tilde{P}_{\sigma, \bar\sigma},\tilde{P}_{g ,\sigma}, \tilde{P}_{e, \sigma},\tilde{P}_{\sigma, e} \right)^T$,
the reduced $10\times 10$ matrix ${\bf M}$ contains the transition rates $\Gamma_{|\alpha,\beta\rangle \to |\alpha',\beta'\rangle}$ listed in Table \ref{tab2}. The diagonal elements are $-\tilde{\Gamma}_{\alpha\beta}$, where 
the respective outgoing rates are given by
\begin{eqnarray*}
\tilde \Gamma_{g,g} & = & \Gamma_{-1,1} + \Gamma_{-2,2} + 2 \left( \Gamma_{-1,2} + \Gamma^{\rm in}_1 + \Gamma^{\rm in}_2 \right),\label{eq1} \\
\tilde \Gamma_{e,g} & = & \Gamma_{1,-1} + \Gamma_{-2,2} + 2 \left( \Gamma_{1,2} + \Gamma^{\rm in}_{-1} + \Gamma^{\rm in}_2 \right), \label{eq2} \\
\tilde \Gamma_{g,e} & = & \Gamma_{-1,1} + \Gamma_{2,-2} + 2 \left( \Gamma_{-1,-2} + \Gamma^{\rm in}_1 + \Gamma^{\rm in}_{-2}  \right),\label{eq3} \\
\tilde \Gamma_{e,e} & = & \Gamma_{1,-1} + \Gamma_{2,-2} + 2 \left( \Gamma_{1,-2} + \Gamma^{\rm in}_{-1} + \Gamma^{\rm in}_{-2} \right),\label{eq4} \\
\tilde \Gamma_{\sigma, \sigma} & = & \sum_{\lambda = \pm 1, \pm 2} \Gamma^{\rm in}_\lambda, \label{eq5} \\
\tilde \Gamma_{\sigma, \bar\sigma} & = & \Gamma_{1,-2} + \Gamma_{1,2} + \Gamma_{-1,-2} + \Gamma_{-1,2} +
\sum_{\lambda = \pm 1, \pm 2} \Gamma^{\rm in}_\lambda,\label{eq6}\\
\tilde \Gamma_{g,\sigma} & = & \Gamma_{-1,-2} + \Gamma_{-1,1} + \Gamma_{-1,2} +
2 \Gamma^{\rm in}_1 + \Gamma^{\rm in}_2 + \Gamma^{\rm in}_{-2}, \label{eq7} \\
\tilde \Gamma_{\sigma ,g} & = & \Gamma_{1,2} + \Gamma_{-1,2} + \Gamma_{-2,2} +
2 \Gamma^{\rm in}_2 + \Gamma^{\rm in}_1 + \Gamma^{\rm in}_{-1}, \label{eq8} \end{eqnarray*}
\begin{eqnarray*}
\tilde \Gamma_{e ,\sigma} & = & \Gamma_{1,-1} + \Gamma_{1,-2} + \Gamma_{12} +
2 \Gamma^{\rm in}_{-1} + \Gamma^{\rm in}_2 + \Gamma^{\rm in}_{-2},\label{eq9} \\
\tilde \Gamma_{\sigma ,e} & = & \Gamma_{1,-2} + \Gamma_{2,-2} + \Gamma_{-1,-2} +
2 \Gamma^{\rm in}_{-2} + \Gamma^{\rm in}_1 + \Gamma^{\rm in}_{-1}.
\label{eq10}
\end{eqnarray*}

\begin{table}
\begin{tabular}{|l||c|c|c|c|}\hline 
                  & $\alpha_1\beta_1$ &  $\alpha_2\beta_2$ & $C_A$            & $C_A^\ast$ \\   \hline \hline
$\alpha_1\beta_1$ & 0                 &  0                 & $ig_d$           & $-ig_d$ \\
$\alpha_2\beta_2$ & 0                 & 0                  & $-ig_d$          & $ig_d$\\
$C_A^{}$          &  $ig_d$            & $-ig_d$              & $-\tilde \Gamma_A$ & 0 \\
$C_A^\ast$        & $-ig_d$             & $i g_d$              & 0                & $-\tilde \Gamma_A$ \\ \hline
\end{tabular}
\caption{\label{tab3} Nonzero matrix elements of the matrix ${\bf N}_d$ in Eq.~\eqref{dotW}. }
\end{table}

In Sec.~\ref{sec3a}, we discuss a simplified approach where an initial condition for Eq.~\eqref{dotP} simulates the effects of a microwave driving pulse.
When driving, e.g., the transition $| \sigma,g \rangle \to | g,\sigma \rangle$ (keeping in mind that states with $\sigma=\pm$ are degenerate), this initial condition is given by
\begin{equation}\label{initcond1}
\tilde{P}_{g,\sigma}(0)=\tilde{P}_{\sigma,g}(\infty)+\tilde{P}_{g,\sigma}(\infty),\quad \tilde{P}_{\sigma,g}(0)=0.
\end{equation}
On the other hand, if the mixed even transition is driven, we only need to exchange the populations of 
the $|g,g\rangle$ and the $|\sigma,\bar\sigma\rangle$ states, 
\begin{equation}\label{initcond}
\tilde{P}_{\sigma,\bar{\sigma}}(0)=P_{g,g}(\infty)+\tilde{P}_{\sigma,\bar{\sigma}}(\infty), \quad P_{g,g}(0)=0 .
\end{equation}

Going beyond the simplified picture, we find in Sec.~\ref{sec3c} that Eq.~\eqref{dotP} has to be
replaced by Eq.~\eqref{dotW} with a matrix ${\bf N}_d$.
The nonzero entries of ${\bf N}_d$ are specified in Table \ref{tab3}. 
For the transitions considered in Sec.~\ref{sec3c}, the coherence decay rates $\tilde \Gamma_A$ are given by
\begin{eqnarray} \nonumber
&&\tilde \Gamma_{A}^{|g,g\rangle\to |\sigma,\bar{\sigma}\rangle} = \frac{1}{2}\Bigl(\Gamma_{1,2}+\Gamma_{-1,2}+\Gamma_{1,-2}+\Gamma_{-1,2}+\Gamma_{2,1} 
\\ \nonumber &&\quad +\Gamma_{-1,1}+\Gamma_{-2,1}+\Gamma_{-2,2}+\Gamma^{\rm out}_{1}+2\Gamma^{\rm out}_{-1}+\Gamma^{\rm out}_{-2}
\\ &&\quad + 2\Gamma^{\rm in}_{2}+\Gamma^{\rm in}_{-2}+\Gamma^{\rm in}_{1} \Bigr),
\label{gammaA}
\\ \nonumber
&&\tilde \Gamma_{A}^{|\sigma,g\rangle\to |g,\sigma\rangle}= \frac{1}{2}\Bigl(\Gamma_{-2,1}+\Gamma_{1,2}+\Gamma_{-2,2}+\Gamma_{-1,1}+\Gamma_{2,1} 
\\ \nonumber &&\quad +\Gamma^{\rm out}_{1}+2\Gamma^{\rm out}_{-1}+\Gamma^{\rm out}_{-2}+ 2\Gamma^{\rm in}_{2}+\Gamma^{\rm in}_{-2}+\Gamma^{\rm in}_{1} \Bigr),\\
 \nonumber &&\tilde \Gamma_{A}^{|e,g\rangle\to |\sigma,\bar{\sigma}\rangle}= \frac{1}{2}\Bigl(3\Gamma_{1,2}+\Gamma_{-1,2}+\Gamma_{1,-2}+\Gamma_{2,1}+\Gamma_{1,-1}
\\ && \quad +\Gamma_{-2,2} \nonumber 
 +2\Gamma^{\rm out}_{1}+\Gamma^{\rm out}_{-1}+\Gamma^{\rm out}_{-2}+ 2\Gamma^{\rm in}_{2}+\Gamma^{\rm in}_{-2}+\Gamma^{\rm in}_{-1} \Bigr).
\end{eqnarray}

\bibliographystyle{aipnum4-1}
\bibliography{suplink}
\end{document}